\documentclass[multphys,vecphys]{svmult}
\usepackage{makeidx}     
\usepackage{graphicx}    
\usepackage{multicol}    

\makeindex             

\newcommand{\EQ}{\begin{equation}}
\newcommand{\EN}{\end{equation}}
\newcommand{\EQA}{\begin{eqnarray}}
\newcommand{\ENA}{\end{eqnarray}}
\newcommand{\eq}[1]{(\ref{#1})}
\newcommand{\EEq}[1]{Equation~(\ref{#1})}
\newcommand{\Eq}[1]{Eq.~(\ref{#1})}
\newcommand{\Eqs}[2]{Eqs~(\ref{#1}) and~(\ref{#2})}

\newcommand{\Sec}[1]{Sect.~\ref{#1}}

\newcommand{\Secss}[2]{Sects~\ref{#1}--\ref{#2}}
\newcommand{\Fig}[1]{Fig.~\ref{#1}}

\newcommand{\Tab}[1]{Table~\ref{#1}}

\newcommand{\bra}[1]{\langle #1\rangle}

\newcommand{\meanEMF}{\overline{\vec{\cal E}}}

\newcommand{\meanFF}{\overline{\mbox{\boldmath ${\cal F}$}} {}}

\newcommand{\meanB}{\overline{B}}

\newcommand{\meanAA}{\overline{\vec{A}}}
\newcommand{\meanBB}{\overline{\vec{B}}}
\newcommand{\meanJJ}{\overline{\vec{J}}}
\newcommand{\meanUU}{\overline{\vec{U}}}
\newcommand{\meanWW}{\overline{\vec{W}}}

\newcommand{\meanEE}{\overline{\vec{E}}}
{}
{}
{}
{}
{}
{}
\newcommand{\nnn}{\hat{\mbox{\boldmath $n$}} {}}

\newcommand{\ddelta}{{\vec{\delta}}}
\newcommand{\xx}{{\vec{x}}}
\newcommand{\UU}{{\vec{U}}}
\newcommand{\uu}{{\vec{u}}}
\newcommand{\BB}{{\vec{B}}}
\newcommand{\JJ}{{\vec{J}}}
\newcommand{\jj}{{\vec{j}}}
\newcommand{\AAA}{{\vec{A}}}
\newcommand{\aaaa}{{\vec{a}}}
\newcommand{\bb}{{\vec{b}}}
\newcommand{\cc}{{\vec{c}}}

\newcommand{\ee}{\mbox{\boldmath $e$} {}}
\newcommand{\ff}{\mbox{\boldmath $f$} {}}

\newcommand{\EE}{{\vec{E}}}

\newcommand{\kk}{{\vec{k}}}

\newcommand{\grav}{\mbox{\boldmath $g$} {}}
\newcommand{\nab}{\vec{\nabla}}

\newcommand{\oo}{\vec{\omega}}

\newcommand{\SSSS}{\mbox{\boldmath ${\sf S}$} {}}

\newcommand{\emf}{\mbox{\boldmath ${\cal E}$} {}}

\newcommand{\e}{{\rm e}}
\newcommand{\ii}{{\rm i}}

\newcommand{\curl}{{\rm curl} \, {}}

\newcommand{\dd}{{\rm d} {}}
\newcommand{\const}{{\rm const}  {}}

\def\ga{\mathrel{\mathchoice {\vcenter{\offinterlineskip\halign{\hfil
$\displaystyle##$\hfil\cr>\cr\sim\cr}}}
{\vcenter{\offinterlineskip\halign{\hfil$\textstyle##$\hfil\cr>\cr\sim\cr}}}
{\vcenter{\offinterlineskip\halign{\hfil$\scriptstyle##$\hfil\cr>\cr\sim\cr}}}
{\vcenter{\offinterlineskip\halign{\hfil$\scriptscriptstyle##$\hfil\cr>\cr\sim\cr}}}}}
\def\half{{\textstyle{1\over2}}}

\def\onethird{{\textstyle{1\over3}}}

\newcommand{\yjgr}[3]{: #1, {JGR} {#2}, #3}

\newcommand{\ysph}[3]{: #1, {Sol. Phys.} {#2}, #3}
\newcommand{\yapj}[3]{: #1, {ApJ} {#2}, #3}
\newcommand{\yapjl}[3]{: #1, {ApJ} {#2}, #3}
\newcommand{\yapjs}[3]{: #1, {ApJS} {#2}, #3}
\newcommand{\yan}[3]{: #1, {AN} {#2}, #3}

\newcommand{\yana}[3]{: #1, {A\&A} {#2}, #3}

\newcommand{\ygafd}[3]{: #1, {Geophys. Astrophys. Fluid Dyn.} {#2}, #3}

\newcommand{\yjfm}[3]{: #1, {JFM} {#2}, #3}
\newcommand{\ypf}[3]{: #1, {Phys. Fluids} {#2}, #3}
\newcommand{\ypp}[3]{: #1, {Phys. Plasmas} {#2}, #3}

\newcommand{\yjetp}[3]{: #1, {Sov. Phys. JETP} {#2}, #3}

\newcommand{\yannr}[3]{: #1, {ARA\&A} {#2}, #3}

\newcommand{\yprl}[3]{: #1, {Phys. Rev. Lett.} {#2}, #3}
\newcommand{\ypre}[3]{: #1, {Phys. Rev. E} {#2}, #3}
\newcommand{\yphl}[3]{: #1, {Phys. Lett.} {#2}, #3}

\newcommand{\ymn}[3]{: #1, {MNRAS} {#2}, #3}

\newcommand{\ypr}[3]{: #1, {Phys. Rev.} {#2}, #3}

\newcommand{\yjour}[4]{: #1, {#2} {#3}, #4}

\newcommand{\ybook}[3]{: #1, {#2} (#3)}
\newcommand{\yproc}[5]{: #1, in {\it #3}, ed. #4 (#5), p.\ #2}
\newcommand{\pproc}[4]{: #1, in {\it #2}, ed. #3 (#4) (in press)}

\newcommand{\sjour}[2]{: #1, {#2}}
\begin{document}

\title*{Importance of magnetic helicity in dynamos}

\author{Axel Brandenburg}
\institute{Nordita, Blegdamsvej 17, DK-2100 Copenhagen \O, Denmark
\texttt{brandenb@nordita.dk}}
%
%
\maketitle

Magnetic helicity is nearly conserved and its evolution equation
provides a dynamical feedback on the alpha effect that is distinct
from the conventional algebraic alpha quenching.
The seriousness of this dynamical alpha quenching is particularly
evident in the case of closed or periodic boxes.
The explicit connection with catastrophic alpha quenching is reviewed
and the alleviating effects of magnetic and current helicity fluxes
are discussed.

\section{Introduction}

Let us begin by defining dynamos and helicity.
Dynamos are a class of velocity fields that allow a weak seed magnetic
field to be amplified until some saturation process sets in.
Mathematically, this is described by exponentially growing solutions
of the induction equation.
Simulations have shown that any sufficiently complex flow field
can act as a dynamo if the resistivity is below a certain threshold.
It is in principle not even necessary that the flow is three-dimensional,
only the magnetic field must be three-dimensional because otherwise
one of several antidynamo theorems apply (Cowling 1934, Zeldovich 1957).

Helicity, on the other hand, quantifies the swirl in a vector field.
There is kinetic helicity, which describes the degree to which vortex
lines follow a screw-like pattern, and it is positive for right-handed
screws.
Examples of helical flows are the highs and lows on the weather map.
For both highs and lows the kinetic helicity has the same sign
and is negative (positive) in the northern (southern) hemisphere.
For example, in an atmospheric low, air flows inward, i.e.\ toward the core of
the vortex, and down to the bottom of the atmosphere, but the Coriolis
force makes it spin anti-clockwise, causing left-handed spiraling motions and
hence negative helicity.

A connection between helicity and dynamos has been established already
quite some time ago when Steenbeck et al.\ (1966) calculated
the now famous $\alpha$ effect in mean
field dynamo theory and explained its connection with kinetic helicity.
In this paper we are not so much concerned with kinetic helicity,
but mostly with the magnetic and current helicities.
Quantifying the swirl of magnetic field lines has diagnostic significance,
because magnetic helicity is a topological invariant of the ideal
(non-resistive) equations.
Especially in the solar community the {\it diagnostic} properties of magnetic
helicity have been exploited extensively over the past decade.
However, the use of magnetic helicity as a {\it prognostic} quantity for
understanding the governing nonlinearity of $\alpha$ effect dynamos has
only recently been noted in connection with the magnetic helicity
constraint (Brandenburg 2001, hereafter referred to as B01).

We should emphasize from the beginning that dynamos do not have
to have helicity.
The small scale dynamo of Kazantsev (1968) is an example of a dynamo
that works even without helicity.
Nonhelical dynamos are generally harder to excite than helical dynamos,
but both can generate
fields of appreciable strength if the magnetic Reynolds number is large.
The stretch-twist-fold dynamo also operates with twist (as the name
suggests!), but the orientation of twist can be random, so the net
helicity can be zero.
Simulations have shown that even with zero helicity density everywhere,
dynamos can work (Hughes et al.\ 1996).

It is also possible to generate magnetic fields of large scale once there
is strong shear, even if there is no helicity (Vishniac \& Brandenburg 1997).
This case is very much a topic of current research.
One of the possibilities is is the so-called shear-current effect
(Rogachevskii \& Kleeorin 2003, 2004), but such dynamos still produce
helical large scale magnetic fields.
There is also the possibility of an intrinsically nonlinear dynamo
operating with magnetic helicity flux alone (Vishniac \& Cho 2001).
Thus, it is not necessarily clear that large scale dynamos have to
work with kinetic helicity and the corresponding $\alpha$ effect.
However, there is as yet no convincing example of a dynamo without the
involvement of kinetic helicity that generates large scale magnetic
fields with a degree of coherence that is similar to that observed in
stars and in galaxies, e.g.\ cyclically migrating magnetic fields in the
sun and grand magnetic spirals in some nearby galaxies.
Such fields can potentially be generated by dynamos with an $\alpha$
effect, as has been shown in many papers over the past 40 years; see
Chapters 2, 4, and 10.

There is however a major problem with $\alpha$ effect dynamos;
see Brandenburg (2003), Brandenburg \& Subramanian (2004) for
recent reviews on the issue.
The degree of severity depends on the nature of the problem.
It is most severe in the case of a homogeneous $\alpha$ effect in a
periodic box, which is also when the problem shows up most pronouncedly.
Cattaneo \& Hughes (1996) found that the $\alpha$ effect is quenched
to resistively small values once the mean field becomes a fraction of
the equipartition field strength.
In response to such difficulties three different approaches have been
pursued.
The most practical one is to simply ignore the problem and the proceed
as if we can still use the $\alpha$ effect with a quenching that only
sets in at equipartition field strengths.
This can partially be justified by the apparent success in applying
this theory; see the recent reviews by Beck et al.\ (1996),
Kulsrud (1999), and Widrow (2002).
The second approach is to resort to direct three dimensional simulations
of the turbulence in such astrophysical bodies.
In the solar community this approach has been pioneered by Gilman (1983)
and Glatzmaier (1985), and more recently by Brun et al.\ (2004).
The third approach is a combination of the first two, i.e.\ to use
direct simulations of problems where mean field theory should give a
definitive answer.
This is also the approach taken in the present work.
The hope is ultimately to find guidance toward a revised mean field
theory and to test it quantitatively.
A lot of progress has already been made which led to the suggestion that
only a dynamical (i.e.\ explicitly time dependent) theory of $\alpha$
quenching is compatible with the simulation results.
In the present paper we review some of the simulations that have
led to this revised understanding of mean field theory.

The dynamical quenching theory is now quite successful in reproducing
the results from simulations in a closed or periodic box with and
without shear.
In these cases super-equipartition fields are possible, but only after
a resistively long time scale.
In the case of an open box without shear the dynamical quenching theory
is also successful in reproducing the results of simulations, but here
the root mean field strength decreases with increasing magnetic Reynolds
number, suggesting that such a dynamo is unimportant for astrophysical
applications.
Open boxes with shear appear now quite promising, but the theory is
still incomplete and, not surprisingly, there are discrepancies
between simulations and theory.
In fact, it is quite possible that it is not even the $\alpha$ effect that
is important for large scale field regeneration.  Alternatives include
the shear-current effect of Rogachevskii \& Kleeorin (2003, 2004) and
the Vishniac \& Cho (2001) magnetic and current helicity fluxes.
In both cases strong helicity fluxes are predicted by the theory and
such fluxes are certainly also confirmed observationally for the sun
(Berger \& Ruzmaikin 2000, DeVore 2000, Chae 2000, Low 2001).
For the galaxy the issue of magnetic helicity is still very much in its
infancy, but some first attempts in this direction are already being
discussed (Shukurov 2004).

\section{Dynamos in a periodic box}

To avoid the impression that all dynamos have to have helicity, we begin
by commenting explicitly on dynamos that do not have net kinetic helicity,
i.e.\ $|\bra{\oo\cdot\uu}|/(k_{\rm f}\bra{\uu^2})\ll1$, where $k_{\rm f}$
is the wavenumber of the forcing (corresponding to the energy carrying
scale).
Unless the flow also possesses some large scale shear flow (discussed
separately in \Sec{OpenSurface} below), such dynamos are referred to as
small scale dynamos.
The statement made in the introduction that {\it any} sufficiently
complex flow field can act as a dynamo is really only based on experience,
and the statement may need to be qualified for small scale dynamos.
Indeed, whether or not turbulent small scale dynamos work in stars where
the magnetic Prandtl numbers are small ($\mbox{Pr}_{\rm M}\approx10^{-4}$)
is unclear (Schekochihin et al.\ 2004, Boldyrev \& Cattaneo 2004).
Simulations suggest that the critical magnetic Reynolds numbers
increase with decreasing magnetic Prandtl number like
$R_{\rm m,crit}\approx35\mbox{Pr}_{\rm M}$ (Haugen et al.\ 2004).

Throughout the rest of this review, we want to focus attention on large
scale dynamos.
This is where magnetic helicity plays an important role.
Before we explain why in a periodic box nonlinear dynamos operate only on a
resistively slow time scale, it may be useful to illustrate the problem
with some numerical facts.

\begin{figure}[t!]\begin{center}
\includegraphics[width=.99\textwidth]{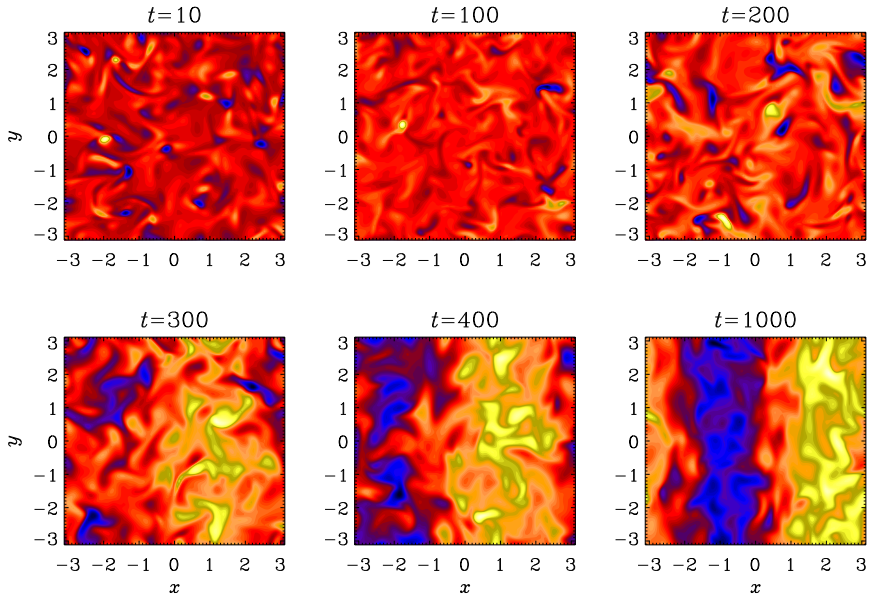}
\end{center}\caption[]{
Cross-sections of $B_x(x,y,0)$ for Run~3 of B01
at different times showing the gradual build-up of
the large scale magnetic field after $t=300$.
The diffusive time scale for this run is $(\eta k_1^2)^{-1}=500$.
Dark (light) corresponds
to negative (positive) values. Each image is scaled with respect to its
min and max values.
The final state corresponds to the second eigenfunction given in
\Eq{BeltramiEigenfunctions}, but with some smaller scale turbulence
superimposed.
[Adapted from Brandenburg \& Subramanian (2004).]
}\label{pimages_hor}\end{figure}

\begin{figure}[t!]\begin{center}
\includegraphics[width=.99\textwidth]{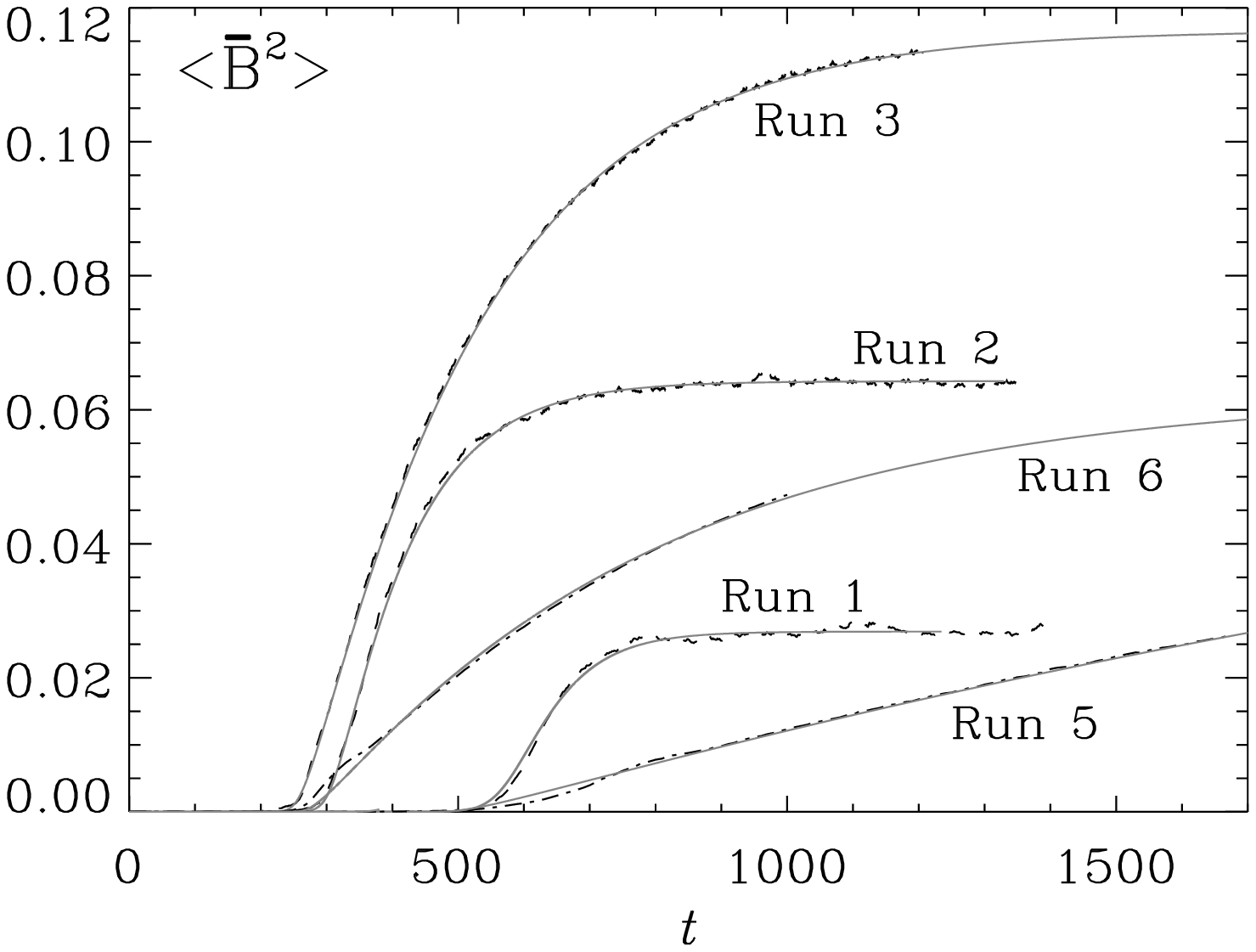}
\end{center}\caption[]{
Evolution of $\bra{\meanBB^2}$ for Runs~1--3 and 5, 6 (dashed lines).
The magnetic Reynolds numbers are
$R_{\rm m}\equiv u_{\rm rms}/\eta k_{\rm f}=2.4$, 6, 18, 100, and 16,
respectively; see B01.
The solid lines denote the solution of the associated mean field dynamo
problem where {\it both} $\alpha$ and and turbulent diffusivity $\eta_{\rm t}$
are quenched in a magnetic Reynolds number dependent fashion.
[Adapted from B01.]
}\label{pjbm_decay_nfit_alpha}\end{figure}

In the simulations of B01 the flow was forced at an intermediate
wavenumber, $k\approx k_{\rm f}=5$, while the smallest wavenumber in the
computational domain corresponds to $k=k_1=1$. 
The kinetic energy spectrum peaks at $k\approx k_{\rm f}$, which is
therefore also the wavenumber of the energy carrying scale. 
The turbulence is nearly fully helical with
$\bra{\oo\cdot\uu}/(k_{\rm f}\bra{\uu^2})\approx0.7...0.9$.
The initial field is random, but as time goes on it develops a large scale
component at wavenumber $k\approx k_1=1$; see \Fig{pimages_hor}.
In \Fig{pjbm_decay_nfit_alpha} we show the evolution of the magnetic energy
of the mean field from the same simulation.\footnote{Here the time unit is
$[t]=(c_{\rm s}k_1)^{-1}$, where $c_{\rm s}$ is the isothermal speed
of sound, and the magnetic field is measured in units of
$[B]=\sqrt{\mu_0\rho_0}c_{\rm s}$.}
Here the mean field is defined as two-dimensional averages over
planes perpendicular to the direction in which the mean field varies.
There are of course three such directions, but there is usually only
one direction for which there is a significant mean field.

While the saturation field strength increases with increasing magnetic
Reynolds number, the time scale on which this nonlinear dynamo saturates
increases too.
To avoid misunderstandings, it is important to emphasize that
this result applies only when we are in the {\it nonlinear regime}
and when the flows are {\it helical}.

In turbulence one is used to situations where the microscopic values of
viscosity $\nu$ and magnetic diffusivity $\eta$ do not matter in the sense
that, for almost all practical purposes, they are superseded by turbulent
effective values, $\nu_{\rm t}$ and $\eta_{\rm t}$, respectively.
This is because in turbulence there is spectral energy all the way down
to the viscous/resistive length scale, $(\eta\tau)^{1/2}$, where $\tau$
is the turnover time.\footnote{The turnover time at the wavenumber $k$
is $(u_kk)^{-1}$. Using Kolmogorov scaling, $u_k\sim k^{-1/3}$, one finds
the familiar formula $k_\eta=k_{\rm f} R_{\rm m}^{3/4}$, where $k_{\rm f}$
is the wavenumber of the energy carrying eddies.}
Thus, even when $\nu$ is very small, the rate of viscous dissipation,
$\bra{2\nu\rho \SSSS^2}$, is in general finite ($\SSSS$ is the trace-less
rate of strain tensor).
Likewise, even when $\eta$ is very small, the rate of Joule dissipation,
$\eta\mu_0\bra{\JJ^2}$, is in general finite ($\mu_0$ is the magnetic
permeability).
This is because the current density diverges with decreasing $\eta$ like
$|\JJ|\sim\eta^{-1/2}$, so the energy dissipation stays finite and
asymptoticly independent of how small $\eta$ is.
The trouble is that the value of magnetic helicity dissipation
is proportional to $\eta\bra{\JJ\cdot\BB}$ (see below), and in the limit
$\eta\to0$ we have $\eta\bra{\JJ\cdot\BB}\to\eta^{1/2}\to0$, so resistive
magnetic helicity dissipation becomes impossible in the limit of large
$R_{\rm m}$.
In the following section we derive and discuss the evolution equation
for magnetic helicity.

\section{Magnetic helicity evolution}

\subsection{The two-scale property of helical turbulence}

Usually in mean field dynamo theory one talks about the two-scale
{\it assumption} made in order to derive the mean field equations
(e.g.\ Moffatt 1978, Krause \& R\"adler 1980).
This has to do with the fact that higher order derivatives in the mean
field equation can only be neglected when the mean field is sufficiently
smooth.
Here, instead, we use the two-scale {\it properties} of helical turbulence
as demonstrated in the previous section.
These properties emerge automatically when the size of the whole body
is at least several times larger than the scale of the turbulent eddies.
As \Fig{pimages_hor} shows explicitly, a large scale field (wavenumber $k_1$)
emerges in addition to the forcing scale (wavenumber $k_{\rm f}\gg k_1$).

In this section we discuss the magnetic helicity equation and use it
together with the two-scale property of helical turbulence
to derive the so-called magnetic
helicity constraint that allows the result of \Fig{pjbm_decay_nfit_alpha}
to be understood quantitatively.

\subsection{Definition of helicity}

The helicity of any solenoidal vector field $\ff$, i.e.\ with
$\nab\cdot\ff=0$, is defined as the volume integral of $\ff$ dotted
with its inverse curl, i.e.\ $\curl^{-1}\ff\equiv\grav$.
As pointed out by Moffatt (1969), the helicity quantifies the
topological linkage between tubes in which $\ff$ is non-vanishing.
In the following the linkage aspect of helicity will not be
utilized, but rather the mathematical evolution equation that
the helicity obeys (see the next section).
However, the calculation of $\grav$ is problematic because it involves
a gauge ambiguity in that the curl of $\grav'=\grav+\nab\phi$ also gives
the same $\ff=\curl\grav'$.

In the special case of periodic boundary conditions or for
$\nnn\cdot\ff=0$ on the boundaries, where $\nnn$ is the normal vector
on the boundary, the helicity is actually gauge-invariant, because
\EQA
\int\ff\cdot\grav'\,\dd V&=&
\int\ff\cdot\grav\,\dd V+\int\ff\cdot\nab\phi\,\dd V
 \nonumber \\
&=& \int\ff\cdot\grav\,\dd V-\int\phi\nab\cdot\ff\,\dd V=
\int\ff\cdot\grav\,\dd V,
\ENA
where we have used $\nab\cdot\ff$.
Since the magnetic field is divergence free, the magnetic helicity,
$\int\BB\cdot\curl^{-1}\BB\,\dd V$ is gauge invariant.
For other boundary conditions this is unfortunately not the case.

For vector fields whose inverse curl is a physically meaningful quantity,
such as the vorticity $\oo$, whose inverse curl is the velocity $\uu$,
the gauge question never arises.
In this and similar cases the helicity density, $\oo\cdot\uu$ in this
case, is physically meaningful.
Other examples are the cross helicity, $\int\BB\cdot\curl^{-1}\oo\,\dd V$,
which describes the linkage between magnetic flux tubes and vortex tubes,
and the current helicity, $\int\JJ\cdot\curl^{-1}\JJ\,\dd V$, which
quantifies the linkage of current tubes.
In these two cases it is natural to use $\curl^{-1}=\BB$ and
$\curl^{-1}\oo=\uu$.
For the magnetic field one can define the magnetic vector potential,
$\curl^{-1}\BB=\AAA$, but $\AAA$ is not a physically meaningful quantity
and hence the magnetic helicity,
\EQ
H=\int_V\AAA\cdot\BB\,\dd V\equiv\bra{\AAA\cdot\BB}\,V
\EN
is gauge-dependent, unless the boundaries of the volume $V$ are periodic
or perfectly conducting.
Here and below, angular brackets denote volume averages.
Occasionally, however, we simply refer to $\bra{\AAA\cdot\BB}$ as the
magnetic helicity, but this is strictly speaking only the magnetic
helicity per unit volume.

In the following section we derive the evolution equation for
$\bra{\AAA\cdot\BB}$ and focus first on the case where the boundary
conditions are indeed periodic, so $\bra{\AAA\cdot\BB}$ is gauge-invariant.

\subsection{Derivation of the magnetic helicity equation}

The homogeneous Maxwell equations are
\EQ
{\partial\BB\over\partial t}=-\nab\times\EE,\quad\nab\cdot\BB=0.
\label{dBdt}
\EN
Expressing this in terms of the magnetic vector potential,
$\AAA$, where $\BB=\nab\times\AAA$, we have
\EQ
{\partial\AAA\over\partial t}=-\EE-\nab\phi,
\label{dAdt}
\EN
where $\phi$ is the scalar potential.
Dotting \Eqs{dBdt}{dAdt} with $\AAA$ and $\BB$, respectively,
and adding them, we have
\EQ
{\partial\over\partial t}(\AAA\cdot\BB)=
-2\EE\cdot\BB-\nab\cdot(\EE\times\AAA+\phi\BB).
\label{dABdt}
\EN
Here, $\AAA\cdot\BB$ is the magnetic helicity density, but since it is not
gauge invariant (see below) it is not a physically meaningful quantity.
After integrating \Eq{dABdt} over a periodic volume, the divergence term does
not contribute.
Furthermore, using Ohm's law, $\EE=-\UU\times\BB+\eta\mu_0\JJ$,
where $\JJ=\nab\times\BB/\mu_0$ is the current density, we have
\EQ
{\dd\over\dd t}\bra{\AAA\cdot\BB}=-2\eta\mu_0\bra{\JJ\cdot\BB},
\label{dABmdt}
\EN
i.e.\ the magnetic helicity, $\bra{\AAA\cdot\BB}$, changes only at a
rate that is proportional to $\eta\bra{\JJ\cdot\BB}$.
(Here and elsewhere, angular brackets denote volume averaging.)
As discussed in the previous section,
this rate converges to zero in the large $R_{\rm m}$ limit.
Here, angular brackets denote volume averages, i.e.\
$\bra{\AAA\cdot\BB}={1\over V}\int_V\AAA\cdot\BB\,\dd V$.

We recall that for periodic boundary conditions,
$\bra{\AAA\cdot\BB}$ is invariant under the
transformation $\AAA\to\AAA'=\AAA+\nab\Lambda$, which does not
change the value of $\BB'=\nab\times\AAA'=\nab\times\AAA=\BB$.
Here, $\Lambda$ is a gauge potential.
Thus, for periodic boundary conditions, $\bra{\AAA\cdot\BB}$ is a physically
meaningful quantity.
The same is also true for perfectly conducting boundaries
(see Brandenburg \& Dobler 2002 for corresponding simulations).
For open boundaries, however, $\int_V\AAA\cdot\BB\,\dd V$ is not gauge
invariant, but one can derive a gauge-invariant relative magnetic helicity
(Berger \& Field 1984).

\subsection{The magnetic helicity constraint}

A very simple argument can be made to explain the saturation level and the
resistively slow saturation behavior observed in \Fig{pjbm_decay_nfit_alpha}.
The only assumption is that the turbulence is helical, i.e.\
$\bra{\oo\cdot\uu}\neq0$, where $\oo$ is the vorticity,
and that this introduces current helicity
$\bra{\jj\cdot\bb}$, at the same scale and of the same sign
as the kinetic helicity.
Here we have split the field into large and small scale
fields, i.e.\ $\BB=\meanBB+\bb$ and hence also $\JJ=\meanJJ+\jj$
and $\AAA=\meanAA+\aaaa$.

The first remarkable thing to note is that, even though we are dealing with
{\it helical} dynamos, there is no net current helicity
in the steady state, i.e.\
\EQ
\bra{\JJ\cdot\BB}=0;
\EN
see \Eq{dABmdt}.
However, using the decomposition into large and small scale fields,
we can write
\EQ
\bra{\JJ\cdot\BB}=\bra{\meanJJ\cdot\meanBB}+\bra{\jj\cdot\bb}=0,
\label{JBJmBmjb}
\EN
so we have
\EQ
\bra{\meanJJ\cdot\meanBB}=-\bra{\jj\cdot\bb}
\label{JmBmjbm}
\EN
in the steady state.
We now introduce the approximations\footnote{Here and elsewhere we use
units where $\mu_0=1$ or, following R.\ Blandford (private communication),
we use units in which pi is one quarter.}
\EQ
\bra{\meanJJ\cdot\meanBB}\approx k_{\rm m}\bra{\meanBB^2}
\quad\mbox{and}\quad
\bra{\jj\cdot\bb}\approx-k_{\rm f}\bra{\bb^2}
\label{JmBm}
\EN
where $k_{\rm m}$ and $k_{\rm f}$ are the typical wavenumbers of the mean
and fluctuating fields, respectively.
These approximations are only valid for fully helical turbulence, but
can easily be generalized to the case of fractional helicity; see
\Sec{Fractional} and Blackman \& Brandenburg (2002, hereafter BB02).
We have furthermore assumed that the sign of the kinetic helicity
is negative, as is the case in the northern hemisphere of the sun,
for example.
(The case of positive kinetic helicity is straightforward; see below.)
The wavenumber $k_{\rm f}$  of the fluctuating field is for all practical
purposes equal to the wavenumber of the forcing function.
(In more general situations, such as convection or shear flow turbulence,
$k_{\rm f}$ would be the wavenumber of the energy carrying eddies.)
We also note that for large values of the magnetic Reynolds number,
$R_{\rm m}$, the $k_{\rm f}$ factor in \Eq{JmBm} gets attenuated by
an $R_{\rm m}^{1/4}$ factor (BB02).
On the other hand, the wavenumber of the mean field is in practice
the wavenumber of the box, i.e.\ $k_{\rm m}=k_1$.
Inserting now \Eq{JmBm} into \eq{JmBmjbm} yields
\EQ
\bra{\meanBB^2}={k_{\rm f}\over k_{\rm m}}\bra{\bb^2},
\label{superequi}
\EN
i.e.\ the energy of the mean field can {\it exceed} the energy
of the fluctuating field -- in contrast to earlier expectations
(e.g.\ Vainshtein \& Cattaneo 1992, Kulsrud \& Andersen 1992,
Gruzinov \& Diamond 1994).
Indeed, in the two-dimensional case there is an exact result due
to Zeldovich (1957),
\EQ
\bra{\meanBB^2}=R_{\rm m}^{-1}\bra{\bb^2}
\quad\mbox{(2-dimensional case)}.
\label{Zeldovich}
\EN
This result has also be derived in three dimensions using first order
smoothing (Krause \& R\"adler 1980), but it is important to realize that
this result can break down in the nonlinear case in three dimensions,
where \Eq{superequi} is in good agreement with the simulations results.
However, the assumption of periodic or closed boundaries is
an essential one.
We return to the more general case in \Secss{Equator}{OpenSurface}.

\begin{figure}[t!]\begin{center}
\includegraphics[width=.99\textwidth]{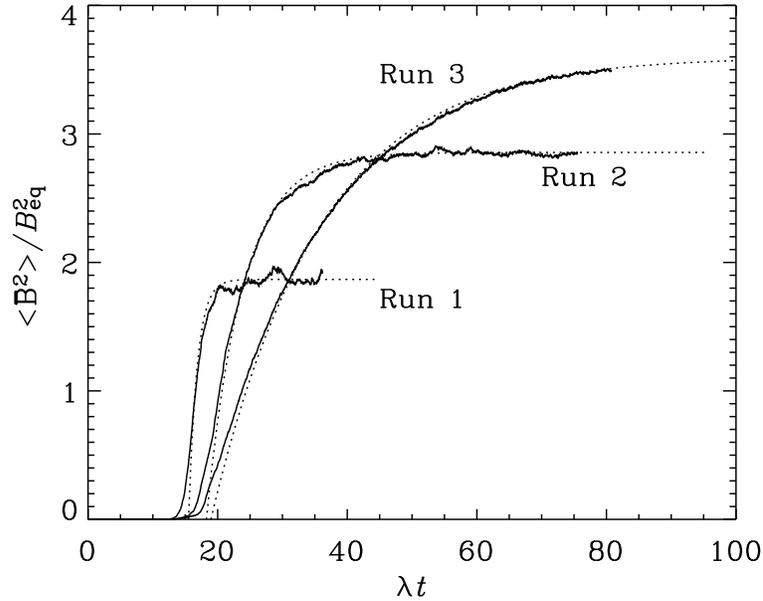}
\end{center}\caption[]{
Late saturation phase of fully helical turbulent dynamos
for three different values of the magnetic Reynolds number:
$R_{\rm m}\equiv u_{\rm rms}/\eta k_{\rm f}=2.4$, 6, and 18 for
Runs~1, 2, and 3 respectively; see B01.
The mean magnetic field, $\meanBB$, is normalized with
respect to the equipartition value,
$B_{\rm eq}=\sqrt{\mu_0\rho_0}u_{\rm rms}$,
and time is normalized with respect to the kinematic
growth rate, $\lambda$.
The dotted lines represent the fit formula \eq{helconstraint}
which tracks the simulation results rather well.
[Adapted from Brandenburg et al.\ (2003).]
}\label{pjbm_decay_nfit}\end{figure}

The time dependence near the saturated state can be approximated
by using
\EQ
\bra{\meanJJ\cdot\meanBB}\approx k_{\rm m}^2\bra{\meanAA\cdot\meanBB}
\quad\mbox{and}\quad
\bra{\jj\cdot\bb}\approx k_{\rm f}^2\bra{\aaaa\cdot\bb}.
\label{JmBmAmBm}
\EN
These equations are still valid in the case of fractional helicity (BB02).
Only the two-scale assumption is required.
Near saturation,
\EQ
|\bra{\meanAA\cdot\meanBB}|=\left({k_{\rm f}\over k_{\rm m}}\right)^2
\bra{\aaaa\cdot\bb},
\EN
i.e.\ $|\bra{\meanAA\cdot\meanBB}|\gg|\bra{\aaaa\cdot\bb}|$ and
so we can neglect $\bra{\aaaa\cdot\bb}$, and the magnetic helicity equation
\eq{dABmdt} becomes therefore an approximate evolution equation for the
magnetic helicity of the mean field,
\EQ
{\partial\over\partial t}\bra{\meanAA\cdot\meanBB}=
-2\eta\mu_0\bra{\meanJJ\cdot\meanBB}
-2\eta\mu_0\bra{\jj\cdot\bb},
\label{dAmBmdt}
\EN
or, by using \Eq{JmBm},
\EQ
k_{\rm m}^{-1}{\partial\over\partial t}\bra{\meanBB^2}=
-2\eta k_{\rm m}\bra{\meanBB^2}+2\eta k_{\rm f}\bra{\bb^2}.
\label{dBm2dt}
\EN
Note the plus sign in front of the $\bra{\bb^2}$ term resulting
from \Eq{JmBm}.
The plus sign leads to growth while the minus sign in front of the
$\bra{\meanBB^2}$ term leads to saturation (but both terms are proportional
to the microscopic value of $\eta$; see below).
Once the small scale field has saturated, which will happen after a few
dynamical time scales such that
$\bra{\bb^2}\approx B_{\rm eq}^2\equiv\mu_0\bra{\rho\uu^2}$,
the large scale field will continue to evolve slowly according to
\EQ
\bra{\meanBB^2}={k_{\rm f}\over k_{\rm m}}\bra{\bb^2}\left[
1-e^{-2\eta k_{\rm m}^2(t-t_{\rm sat})}\right],
\label{helconstraint}
\EN
where $t_{\rm sat}$ is the time at which $\bra{\bb^2}$ has reached
approximate saturation.
In practice, $t_{\rm sat}$ can be determined such that \Eq{helconstraint}
describes the simulation data best.
We refer to \Eq{helconstraint} as the {\it magnetic helicity constraint}.
The agreement between this and the actual simulations
(\Fig{pjbm_decay_nfit}) is quite striking.

The significance of this remarkable and simple equation and the
almost perfect agreement with simulations
is that the constraint can be extrapolated to large
values of $R_{\rm m}$ where it would provide a benchmark, against
which all analytic dynamo theories, when subjected to the same
periodic boundary conditions, should be compared to.
In particular the late saturation behavior should be equally slow.
We return to this in \Sec{ConnectionWithAlphaEffect}.

An important question is whether anything can be learned about stars
and galaxies.
Before this can be addressed, we need to understand the differences
between dynamos in real astrophysical bodies and dynamos in periodic
domains.

\section{What do stars and galaxies do differently?}

We begin with a discussion of fractional helicity, shear and other
effects that cause the magnetic helicity to be reduced.
We then address the possibility of helicity fluxes through boundaries,
which can alleviate the helicity constraint (Blackman \& Field 2000).

\subsection{Fractional helicity}
\label{Fractional}

When the turbulence is no longer fully helical, \Eq{JmBm} is no longer
valid and needs to be generalized to
\EQ
\bra{\meanJJ\cdot\meanBB}=\epsilon_{\rm m}k_{\rm m}\bra{\meanBB^2}
\quad\mbox{and}\quad
\bra{\jj\cdot\bb}=-\epsilon_{\rm f}k_{\rm f}\bra{\bb^2},
\label{JmBmeps}
\EN
where $\epsilon_{\rm m}<1$ and $\epsilon_{\rm f}<1$ are coefficients
denoting the degree to which the mean and fluctuating fields are helical.
\EEq{JmBmAmBm} is still approximately valid in the fractionally helical
case.

Maron \& Blackman (2002) found that there is a certain threshold of
$\epsilon_{\rm f}$ below which the large scale dynamo effect stops
working.
Qualitatively, this could be understood by noting that the large
scale magnetic field comes from the helical part of the flow, so
the velocity field can be though of as having a helical and
a nonhelical component, i.e.\
\begin{equation}
\vec{U}=\vec{U}_{\rm hel}+\vec{U}_{\rm nohel}.
\end{equation}
However, the dynamo effect has to compete with turbulent diffusion which
comes from both the helical and the nonhelical parts of the flow.
Thus, when $|\vec{U}_{\rm nohel}|$ becomes too large compared with
$|\vec{U}_{\rm hel}|$ the large scale dynamo effect will stop working.

Although we have not yet discussed mean field theory we may note that
the value of the threshold can be understood quantitatively
(Brandenburg et al.\ 2002, hereafter BDS02)
and one finds that large scale dynamo action is only possible when
\EQ
\epsilon_{\rm f} > {k_{\rm m}\over k_{\rm f}}
\quad\mbox{(for large scale dynamos)}.
\EN
In many three-dimensional turbulence simulations or in astrophysical
bodies, this threshold criterion may not be satisfied,
and so mean field dynamo of the type described above
($\alpha^2$ dynamo) may not be excited.
If there is shear, however, this criterion will be modified to
\EQ
\epsilon_{\rm f} > \epsilon_{\rm m}{k_{\rm m}\over k_{\rm f}},
\EN
where $\epsilon_{\rm f}$ is the degree to which the large scale field
is helical.
In dynamos with strong shear, $|\epsilon_{\rm f}|$ may be very small,
making mean field dynamo action in fractionally helical flows more likely.
This will be discussed in the next section.

\subsection{Dynamos with shear}

In the presence of shear, the streamwise component of the field can be
amplified by winding up the poloidal (cross-stream) field.
Again, the resulting saturation field strength can be estimated based
on magnetic helicity conservation arguments.

\begin{figure}[t!]\begin{center}
\includegraphics[width=.99\textwidth]{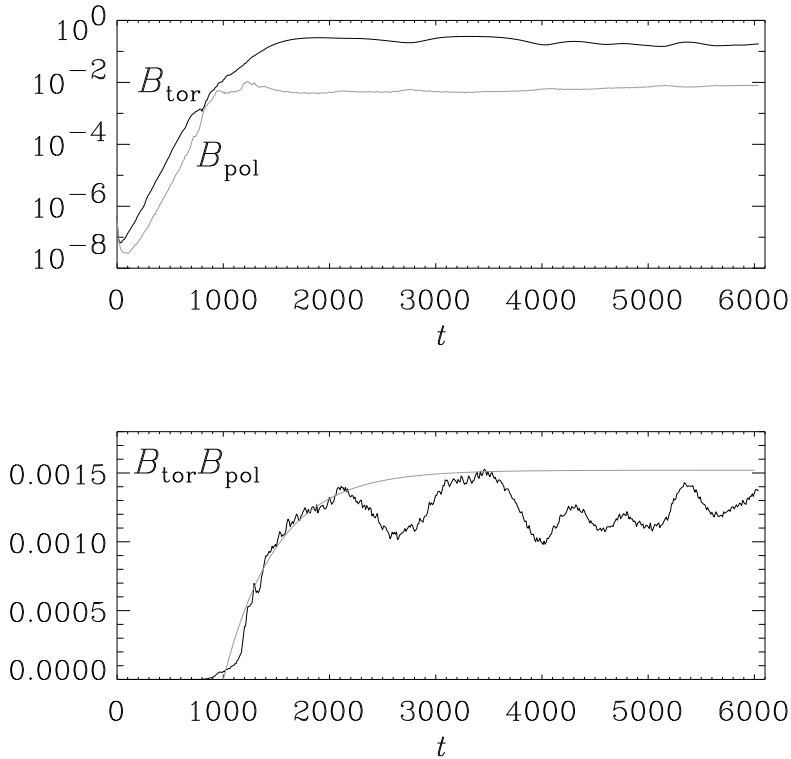}
\end{center}\caption[]{
Growth of poloidal and toroidal magnetic fields on a logarithmic scale
(upper panel), and product of poloidal and toroidal magnetic fields on
a linear scale. For the fit we have used $k_1^2=2$, $B_{\rm eq}=0.035$,
and $\epsilon_0=1.3$.
[Adapted from BBS01.]
}\label{Fpbm1}\end{figure}

Note first of all that for closed or periodic domains, \Eq{JBJmBmjb} is
still valid and therefore $\bra{\meanJJ\cdot\meanBB}=-\bra{\jj\cdot\bb}$
in the steady state.\footnote{This is because in the $\EE\cdot\BB$ term in
the magnetic helicity equation the induction term, $\UU\times\BB$, drops
out after dotting with $\BB$. (For this reason, also ambipolar diffusion
and the Hall effect do not change magnetic helicity conservation.)}
However, while $\bra{\jj\cdot\bb}$ still depends on the helicity of the
small scale field, the corresponding value of $\bra{\meanJJ\cdot\meanBB}$
no longer provides such a stringent bound on $\bra{\meanBB^2}$ as before.
This is because shear can amplify the toroidal field independently of
any magnetic helicity considerations.
The component of $\meanBB^2$ that is amplified by shear is nonhelical
and so we have
\EQ
\epsilon_{\rm m}=
|\bra{\meanJJ\cdot\meanBB}|
\left/\left(k_{\rm m}\bra{\meanBB^2}\right)\ll1\right.
\label{epsmdef}
\EN
(or at least $|\epsilon_{\rm m}|\ll1$ when the helicity of the forcing
is negative and $\epsilon_{\rm m}$ therefore negative).
The value of $\epsilon_{\rm m}$ is proportional to the ratio of
poloidal to toroidal field
\EQ
\epsilon_{\rm m}\approx\pm
2\bra{B_{\rm pol}^2}^{1/2}/\bra{B_{\rm tor}^2}^{1/2},
\EN
where the numerical pre-factor can be different for different
examples.\footnote{Take as an example
$\meanBB(z)=(\meanB_{\rm pol},\meanB_{\rm tor},0)^T
=(\epsilon\cos k_1z,\sin k_1z,0)^T$ for $\epsilon\ll1$,
so $\bra{\meanBB^2}\approx1/2$ and
$\bra{\meanJJ\cdot\meanBB}\approx\epsilon k_1$ and
therefore $\epsilon_{\rm m}=2\epsilon$.}
With these preparations the magnetic helicity constraint can be
generalized to
\EQ
2B_{\rm pol}^{\rm rms}B_{\rm tor}^{\rm rms}\approx
{k_{\rm f}\over k_{\rm m}}\bra{\bb^2}\left[
1-e^{-2\eta k_{\rm m}^2(t-t_{\rm sat})}\right].
\label{helconstraintShear}
\EN
This form of the constraint was proposed and confirmed using three-dimensional
simulations of forced helical turbulence with large scale shear
(Brandenburg et al.\ 2001, hereafter BBS01); see also \Fig{Fpbm1}.

The main conclusion to be drawn from this is that the magnetic helicity
constraint is still valid in the presence of shear, i.e.\ the timescale
of saturation is still controlled by the microscopic magnetic diffusivity.
The only difference is that stronger field strengths are now possible.

Another interesting aspect is that dynamos with shear allow for
oscillatory solutions of the magnetic field.
This is expected from mean field dynamo theory
(Steenbeck \& Krause 1969a,b), but it is also borne out by simulations
(BBS01).
The main result is that the resulting cycle frequency seems to scale
with the microscopic magnetic diffusivity, not the turbulent magnetic
diffusivity.
This confirms again that in a closed domain the magnetic helicity
constraint plays a crucial role in controlling the timescale of
nonlinear dynamos.

\subsection{Hall effect dynamos}

In recent years the importance of the Hall effect has been emphasized
by a number of groups, especially in applications to protostellar
accretion discs (Balbus \& Terquem 2001).
The hall effect can lead to strong nonlinear steepening
of field gradients (Vainshtein et al.\ 2000), and is therefore
important for fast reconnection (e.g.\ Rogers et al.\ 2001),
which in turn is relevant for neutron stars (Hollerbach \& R\"udiger 2004).
Nevertheless, since magnetic helicity generation (and removal) is proportional
to the dot product of electric and magnetic fields, and since the Hall
current is proportional to $\JJ\times\BB$, the Hall term does not affect
magnetic helicity conservation.
Therefore the resistively limited saturation behavior should not be
affected by the Hall term.
Nevertheless, some degree of extra field amplification of the large scale
field has been reported (Mininni et al.\ 2003), and it will be interesting
to identify exactly the processes that led to this amplification.

\subsection{Magnetic helicity exchange across the equator or with depth}
\label{Equator}

The presence of an equator provides a source of magnetic
helicity exchange between domains of negative helicity in the northern
hemisphere (upper disc plane in an accretion disc) and positive helicity
in the southern hemisphere (lower disc plane).
A similar situation can also arise in convection zones where the
helicity is expected to change with depth (Yoshimura 1975).

\begin{figure}[h!]\centering\includegraphics[width=0.95\textwidth]{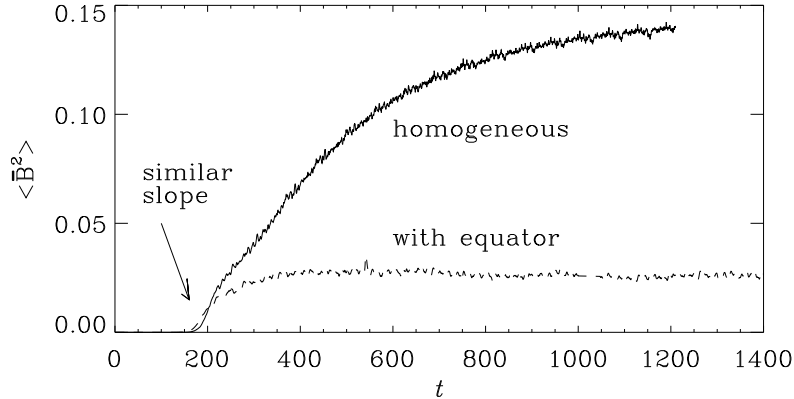}\caption{
Evolution of the magnetic energy for a run with homogeneous forcing
function (solid line) and a forcing function whose helicity varies
sinusoidally throughout the domain (dotted line) simulating the effects
of equators at the two nodes of the sinusoidal helicity profile.
[Adapted from Brandenburg et al.\ (2003).]
}\label{Fpbmean_Run2}\end{figure}

So far, simulations have not yet shown that the losses of small scale
magnetic fields are actually stronger than those of large scales fields.
In \Fig{Fpbmean_Run2} we show the saturation behavior of a system that is periodic,
but the helicity of the forcing is modulated in the $z$-direction
such that the sign of the kinetic helicity changes in the middle.
One can therefore view this system as two subsystems with a boundary
in between. This boundary would correspond to the equator in a star
or the midplane in a disc.
It can also model the change of sign of helicity at some depth in a
convection zone.

As far as the magnetic helicity constraint is concerned, the divergence
term of current helicity flux is likely to be important when there is a
boundary between two domains with different helicities.
Naively, one might expect there to be current helicity fluxes that are
proportional to the current helicity gradient, analogous to Fick's
diffusion law.
These current helicity fluxes should be treated separately for large
and small scale components of the field, so we introduce approximations
to the current helicity fluxes from the mean and fluctuating fields as
\EQ
\meanFF_{\rm m}\approx-\eta_{\rm m}\nab C_{\rm m},\quad
\meanFF_{\rm f}\approx-\eta_{\rm f}\nab C_{\rm f}.
\EN
The rate of magnetic helicity loss is here proportional to some turbulent
diffusivity coefficient, $\eta_{\rm m}$ or $\eta_{\rm f}$ for the losses
from mean or fluctuating parts, respectively.
We assume that the small and large scale fields are
maximally helical (or have known helicity fractions $\epsilon_{\rm m}$
and $\epsilon_{\rm f}$) and have opposite signs of magnetic helicity at small
and large scales.
The details can be found in BDS02 and Blackman \& Brandenburg (2003).
The strength of this approach is that
it is quite independent of mean field theory.

We proceed analogously to the derivation of \Eq{helconstraint}
where we used the
magnetic helicity equation \eq{dABmdt} for a closed
domain to estimate the time derivative of the magnetic helicity of
the mean field by
neglecting the time derivative of the fluctuating field.
This is a good approximation after the fluctuating field has reached
saturation, i.e.\ $t>t_{\rm sat}$.
Thus, we have
\EQ
k_{\rm m}^{-1}{\partial\over\partial t}\meanBB^2=
-2\eta_{\rm m} k_{\rm m}\meanBB^2+2\eta_{\rm f}k_{\rm f}\overline{\bb^2},
\label{dBm2dtflux}
\EN
where $\eta_{\rm m}=\eta_{\rm f}=\eta$ corresponds to the case
of a closed domain.
Note also that we have here ignored the volume integration,
so we are dealing with horizontal averages that depend still on
height and on time.

After the time when the small scale magnetic field saturates, i.e.\
when $t>t_{\rm sat}$, we have $\bra{\bb^2}\approx\mbox{constant}$.
After that time, Eq.~(\ref{dBm2dtflux}) can be solved to give
\begin{equation}
\bra{\meanBB^2}=\bra{\bb^2}{\eta_{\rm f}k_{\rm f}\over\eta_{\rm m}k_{\rm m}}
\left[1-\e^{-2\eta_{\rm m}k_{\rm m}^2(t-t_{\rm sat})}\right],
\quad\mbox{for $t>t_{\rm sat}$}.
\label{solution_pheno}
\end{equation}
This equation demonstrates three remarkable properties
(Brandenburg et al.\ 2003, Brandenburg \& Subramanian 2004):
\begin{itemize}
\item{}
Large scale helicity losses are needed ($\eta_{\rm m}>\eta$)
to shorten the typical time scale. This is required to prevent
resistively long cycle periods.
\item{}
However, the saturation amplitude is proportional to
$\eta_{\rm f}/\eta_{\rm m}$, so the large scale field becomes weaker
as $\eta_{\rm m}$ is increased. Thus,
\item{}
also small scale losses are needed to prevent the saturation amplitude
from becoming too small.
\end{itemize}

Future work can hopefully verify that these conditions are indeed
obeyed by a working large scale dynamo.
Simulations without shear have been unsuccessful to demonstrate that
small scale losses are important (Brandenburg \& Dobler 2001), but new
simulations with shear now begin to show significant small scale losses
of current helicity, an enhanced $\alpha$ effect
(Brandenburg \& Sandin 2004), and strong large scale dynamo action
(see below).

\subsection{Open surfaces and shear}
\label{OpenSurface}

\begin{figure}[t!]\centering\includegraphics[width=0.95\textwidth]{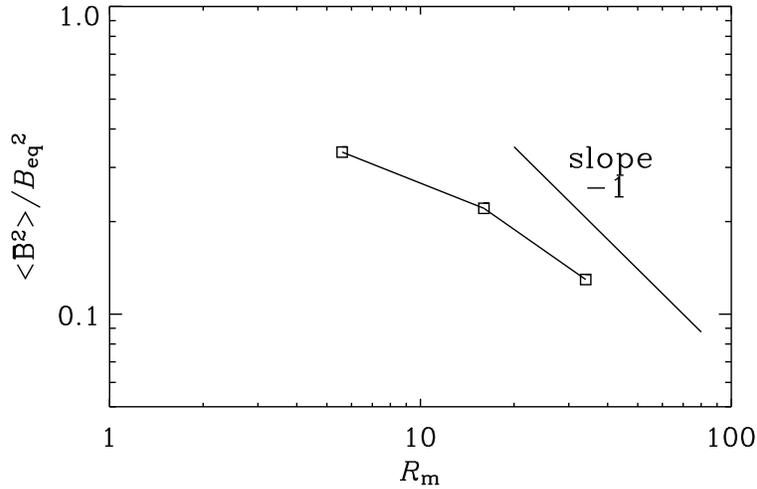}\caption{
Dependence of the energy of the mean magnetic field on the magnetic
Reynolds number for a run with open boundary conditions and no shear.
}\label{Fpbmrm}\end{figure}

The presence of an outer surface is in many respects similar to the
presence of an equator.
In both cases one expects magnetic and current helicity fluxes via
the divergence term.
A particularly instructive system is helical turbulence in an
infinitely extended horizontal slab with stress-free boundary
conditions and a vertical field condition, i.e.\
\EQ
u_{x,z}=u_{y,z}=u_{z}=B_x=B_y=0.
\EN
Such simulations have been performed by Brandenburg \& Dobler (2001) who
found that a mean magnetic field is generated, similar to the case with
periodic boundary conditions, but that the energy of the mean magnetic
field, $\bra{\meanBB^2}$, decreases with magnetic Reynolds number.
Nevertheless, the energy of the total magnetic field, $\bra{\BB^2}$,
does not decrease with increasing magnetic Reynolds number.
Although they found that $\bra{\meanBB^2}$ decreases only like $R_{\rm
m}^{-1/2}$, new simulations confirm that a proper scaling regime has
not yet been reached and that the current data may well be compatible
with an $R_{\rm m}^{-1}$ dependence; see \Fig{Fpbmrm}.

Clearly, an asymptotic decrease of the mean magnetic field must mean
that the small scale dynamo does not work with such boundary conditions.
Thus, the anticipated advantages of open boundary conditions are not
borne out by this type of simulations.

\begin{figure}[t!]
\centering\includegraphics[width=0.95\textwidth]{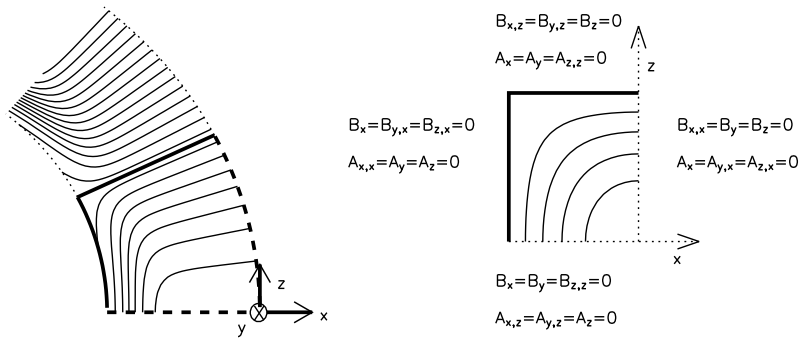}\caption{
Left: A sketch of the solar angular velocity at low latitudes with
spoke-like contours in the bulk of the convection zone merging gradually
into uniform rotation in the radiative interior.
The low latitude region, modeled in this paper, is indicated by thick
lines.
Right: Differential rotation in our cartesian model, with the equator
being at the bottom, the surface to the right, the bottom of the
convection zone to the left and mid-latitudes at the top.
[Adapted from Brandenburg \& Sandin (2004).]
}\label{sketch1}\end{figure}

At this point we can mention some new simulations in a cartesian
domain where differential rotation has been modeled as a
region of the convection zone without explicitly allowing for
convection; see \Fig{sketch1}.
Instead, an external forcing term has been applied that also
drives the differential rotation.
(Studies of the $\alpha$ effect have already been published;
see \Sec{alphaOPEN} for details of the simulations and
\Sec{ConnectionWithAlphaEffect} for a discussion of the direct
correspondence between the helicity constraint and the so-called
catastrophic $\alpha$ quenching.)
Here we briefly report on recent explicit dynamo simulations that have been 
carried out in this geometry.

\begin{figure}[t!]
\centering\includegraphics[width=0.95\textwidth]{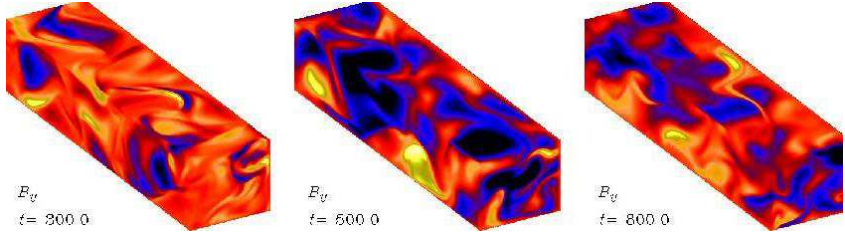}\caption{
Visualization of the toroidal magnetic field during three different times
during the growth and saturation for the run without kinetic helicity.
}\label{diffrot_small}\end{figure}

The size of the computational domain is $\half\pi\times2\pi\times\half\pi$
and the numerical resolution is $128\times512\times128$ meshpoints.
The magnetic Reynolds number based on the forcing wavenumber and the
turbulent flow is around 80 and shear flow velocity exceeds the rms
turbulent velocity by a factor of about 5.
We have carried out experiments with no helicity in the forcing
(labeled by $\alpha=0$), as well as positive and negative helicity
in the forcing (labeled by $\alpha<0$ and $\alpha>0$, respectively);
see \Fig{diffrot_small} for a visualization of the run without
kinetic helicity.
We emphasize that no explicit $\alpha$ effect has been invoked.
The labeling just reflects the fact that, in isotropic turbulence, negative
kinetic helicity (as in the northern hemisphere of a star or the upper
disc plane in galaxies) leads
to a positive $\alpha$ effect, and vice versa.

\begin{figure}[t!]
\centering\includegraphics[width=0.95\textwidth]{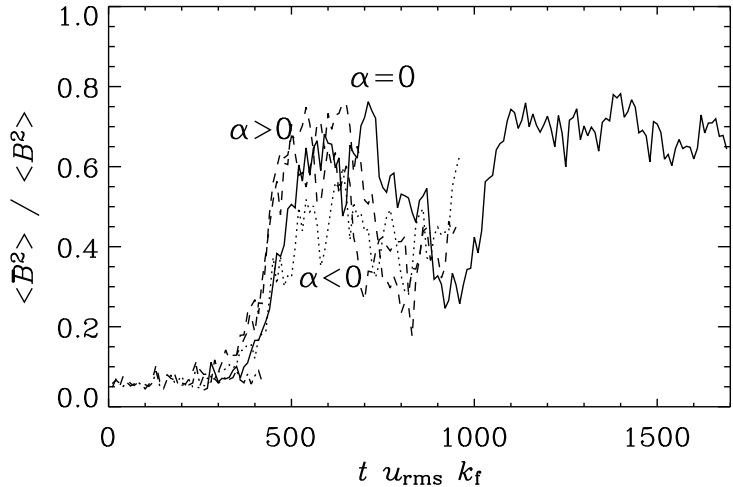}\caption{
Saturation behavior of the ratio $q=\bra{\meanBB^2}/\bra{\BB^2}$
for runs with different kinetic helicity of the flow.
Solid line: zero helicity,
dotted line: positive helicity (opposite to the sun)
dashed line: negative helicity (as in the sun).
}\label{pmean_comp}\end{figure}

We characterize the relative strength of the mean field by the
ratio $q=\bra{\meanBB^2}/\bra{\BB^2}$, where overbars denote an average
in the toroidal ($y$) direction; see \Fig{pmean_comp}.
There are two surprising results emerging from this work.
First, in the presence of shear rather strong mean fields can be
generated, where up to 70\% of the energy can be in the mean field;
see \Fig{pmean_comp}.
Second, even without any kinetic helicity in the flow there is strong
large scale field generation.
Obviously, this cannot be an $\alpha\Omega$ dynamo in the usual sense.
One possibility is the $\ddelta\times\JJ$ effect, which emerged
originally in the presence of the Coriolis force; see R\"adler (1969)
and Krause \& R\"adler (1980).
In the present case with no Coriolis force, however, a $\ddelta\times\JJ$
effect is possible even in the presence of shear alone, because
the vorticity associated with the shear contributes directly to
$\delta\propto\meanWW=\nab\times\meanUU$ (Rogachevskii \& Kleeorin 2003).

There is evidence that the strong dynamo action seen in the
simulations is only possible due to the combined presence of open
boundaries and shear.
This however has so far only been checked explicitly for the $\alpha$
effect that is present when the forcing is helical; see \Sec{alphaOPEN}.
In the case of the solar surface such losses are actually observed to
occur in the form of coronal mass ejections and in active regions.
In the sun, coronal mass ejections are quite vigorous
events that are known to shed large amounts of helical magnetic fields
(Berger \& Ruzmaikin 2000, DeVore 2000, Chae 2000, Low 2001).
This kind of physics is not at all represented by adopting vacuum
or pseudo-vacuum (vertical field) boundary conditions, as was done
in Brandenburg \& Sandin (2004).

\section{Connection with the $\alpha$ effect}
\label{ConnectionWithAlphaEffect}

\subsection{Preliminary considerations}

The $\alpha$ effect formalism provides so far the only workable
mathematical framework for describing the large scale dynamo action seen
in simulations of helically forced turbulence.
(In this section we retain the $\mu_0$ factor.)
The governing equation for the mean magnetic field is
\EQ
{\partial\meanBB\over\partial t}=\nab\times\left(
\meanUU\times\meanBB+\meanEMF-\eta\mu_0\meanJJ\right),
\EN
where $\meanEMF=\overline{\uu\times\bb}$ is the electromotive
force resulting from the $\uu\times\bb$ nonlinearity in the
averaged Ohm's law.
Without mean flow, $\meanUU=0$, and an electromotive force given by a
homogeneous isotropic $\alpha$ effect and turbulent diffusion
$\eta_{\rm t}$, i.e.\
\EQ
\meanEMF=\alpha\meanBB-\eta_{\rm t}\mu_0\meanJJ,
\label{emf_expression}
\EN
we have
\EQ
{\partial\meanBB\over\partial t}=\alpha\nab\times\meanBB
+(\eta+\eta_{\rm t})\nabla^2\meanBB,
\label{constcoeff}
\EN
which has solutions of the form
$\meanBB=\hat{\meanBB}\e^{\ii\kk\cdot\xx+\lambda t}$
with the dispersion relation
\EQ
\lambda_\pm=-\eta_{\rm T} k^2\pm|\alpha k|,
\label{a2disperrel}
\EN
and three possible eigenfunctions (appropriate for the periodic box)
\EQ
\meanBB(\xx)=\pmatrix{\cos k_{\rm m}z\cr\sin k_{\rm m}z\cr 0},\quad
\pmatrix{0\cr\cos k_{\rm m}x\cr\sin k_{\rm m}x},\quad\mbox{or}\quad
\pmatrix{\sin k_{\rm m}y\cr0\cr\cos k_{\rm m}y},
\label{BeltramiEigenfunctions}
\EN
where $k_{\rm m}=k_1=1$.
Obviously, when the coefficients $\alpha$ and
$\eta_{\rm T}\equiv\eta+\eta_{\rm t}$ remain constant, and there is an
exponentially growing solution (for $|\alpha|>\eta_{\rm T}k_1$),
the solution must eventually grow beyond any bound.
At the latest when the magnetic field reaches equipartition with the kinetic
energy, $\alpha$ and $\eta_{\rm t}$ must begin to depend on the
magnetic field itself.
However, the present case is sufficiently simple so that we can continue
to assume that $\meanBB^2$, as well as $\alpha$ and $\eta_{\rm t}$,
are uniform in space and depend only on time.

Comparison with simulations has enabled us to eliminate a large
number of various quenching models where $\alpha=\alpha(\meanBB)$.
The only quenching model that seems reasonably well compatible with
simulations of $\alpha^2$-like dynamo action in a periodic box
{\it without} shear is
\EQ
\alpha={\alpha_0\over1+R_{\rm m}\meanBB^2/B_{\rm eq}^2},\quad
\eta_{\rm t}={\eta_{\rm t0}\over1+R_{\rm m}\meanBB^2/B_{\rm eq}^2}
\quad\mbox{(empirical)},
\label{alpetaCatastrophe}
\EN
see \Fig{pjbm_decay_nfit}.
However, this type of quenching is not fully compatible
with magnetic helicity conservation, as has been shown by
Field \& Blackman (2002).
This will be discussed in the next section.

\subsection{Dynamical $\alpha$ quenching}

The basic idea is that magnetic helicity conservation must be obeyed,
but the presence of an $\alpha$ effect leads to magnetic helicity of
the mean field which has to be balanced by magnetic helicity of
the fluctuating field.
This magnetic helicity of the fluctuating (small scale) field must
be of opposite sign to that of the mean (large scale) field.

We begin with the uncurled mean-field induction equation, written
in the form
\EQ
{\partial\meanAA\over\partial t}=\meanEMF-\eta\mu_0\meanJJ,
\EN
dot it with $\meanBB$, add the result to
$\meanAA\cdot\partial\meanBB/\partial t$, average over the periodic box,
and obtain
\EQ
{\partial\over\partial t}\bra{\meanAA\cdot\meanBB}
=2\bra{\meanEMF\cdot\meanBB}-2\eta\mu_0\bra{\meanJJ\cdot\meanBB}.
\EN
To satisfy the helicity equation for the full field,
$\bra{\AAA\cdot\BB}=\bra{\meanAA\cdot\meanBB}+\bra{\aaaa\cdot\bb}$,
we must have
\EQ
{\partial\over\partial t}\bra{\aaaa\cdot\bb}
=-2\bra{\meanEMF\cdot\meanBB}-2\eta\mu_0\bra{\jj\cdot\bb}.
\label{dabmdt}
\EN
Note the minus sign in front of the $2\bra{\meanEMF\cdot\meanBB}$ term,
indicating once again that the $\alpha$ effect produces magnetic helicity
of opposite sign at the mean and fluctuating fields.
The sum of the two equations yields \Eq{dABmdt}.

The significance of \Eq{dabmdt} is that it contains the $\bra{\jj\cdot\bb}$
term which contributes to the $\alpha$ effect, as was first shown by
Pouquet et al.\ (1976).
Specifically, they found (see also Blackman \& Field 2002)
\EQ
\alpha=\alpha_{\rm K}+\alpha_{\rm M},\quad\mbox{with}\quad
\alpha_{\rm K}=-\onethird\tau\bra{\oo\cdot\uu},\quad
\alpha_{\rm M}=+\onethird\tau\bra{\jj\cdot\bb},
\label{alpPFM}
\EN
where $\tau$ is the correlation time of the turbulence,
$\oo=\nab\times\uu$ is the vorticity, and $\bra{\oo\cdot\uu}$
is the kinematic helicity.

Using $\bra{\jj\cdot\bb}\approx k_{\rm f}^2\bra{\aaaa\cdot\bb}$,
see \Eq{JmBmAmBm}, we can rewrite \Eq{dabmdt} in a form that can
directly be used in mean field calculations:
\EQ 
{\dd\alpha\over\dd t}=-2\eta_{\rm t} k_{\rm f}^2\left(
{\alpha\bra{\meanBB^2}-\eta_{\rm t}\bra{\meanJJ\cdot\meanBB}
\over B_{\rm eq}^2}+{\alpha-\alpha_{\rm K}\over R_{\rm m}}\right),
\label{fullset2}
\EN 
Here we have used $\eta_{\rm t}=\onethird\tau u_{\rm rms}^2$ to eliminate
$\tau$ in favor of $\eta_{\rm t}$ and $B_{\rm eq}^2=\mu_0\rho_0 u_{\rm rms}^2$
to eliminate $u_{\rm rms}^2$ in favor of $B_{\rm eq}^2$.

So, $\alpha$ is no longer just an algebraic function of $\meanBB$,
but it is related to $\meanBB$ via a
{\it dynamical}, explicitly time-dependent equation.
In the context of dynamos in periodic domains, where magnetic helicity
conservation is particularly important, the time dependence of $\alpha$
can hardly be ignored, unless one wants to describe the final stationary
state, which can be at the end of a very slow saturation phase.
However, in order to make contact with earlier work, it is useful
to consider the stationary limit of \Eq{fullset2}, i.e.\ set
$\partial\alpha/\partial t$.

\subsection{Steady limit and its limitations}

In the steady limit the term in brackets in \Eq{fullset2} can be set
to zero, so this equation reduces to
\EQ
R_{\rm m}{\alpha\bra{\meanBB^2}-\eta_{\rm t}\bra{\meanJJ\cdot\meanBB}
\over B_{\rm eq}^2}+\alpha=\alpha_{\rm K}
\quad\mbox{(for $\dd\alpha/\dd t=0$)}.
\label{fullset1steady}
\EN
Solving this equation for $\alpha$ yields (Kleeorin \& Ruzmaikin 1982,
Gruzinov \& Diamond 1994)
\EQ
\alpha={\alpha_{\rm K}
+\eta_{\rm t} R_{\rm m}\bra{\meanJJ\cdot\meanBB}/B_{\rm eq}^2
\over1+R_{\rm m}\bra{\meanBB^2}/B_{\rm eq}^2}
\quad\mbox{(for $\dd\alpha/\dd t=0$)}.
\label{AlphaStationary}
\EN
And, sure enough, for the numerical experiments with an imposed large
scale field over the scale of the box (Cattaneo \& Hughes 1996), where $\meanBB$
is spatially uniform and therefore $\meanJJ=0$, one recovers the
`catastrophic' quenching formula,
\EQ
\alpha={\alpha_{\rm K}\over1+R_{\rm m}\bra{\meanBB^2}/B_{\rm eq}^2}
\quad\mbox{(for $\meanJJ=0$)},
\EN
which implies that $\alpha$ becomes quenched when
$\bra{\meanBB^2}/B_{\rm eq}^2=R_{\rm m}^{-1}\approx10^{-8}$
for the sun, and for even smaller fields in the case of galaxies.

On the other hand, if the mean field
is not imposed, but maintained by dynamo action,
$\meanBB$ cannot be spatially uniform and then $\meanJJ$ is finite.
In the case of a Beltrami field \eq{BeltramiEigenfunctions},
$\bra{\meanJJ\cdot\meanBB}/\bra{\meanBB^2}\equiv\tilde{k}_{\rm m}$
is some effective wavenumber of the large scale field
[$\tilde{k}_{\rm m}=\epsilon_{\rm m}k_{\rm m}$; see \Eq{epsmdef}].
Since $R_{\rm m}$ enters both the numerator and the denominator,
$\alpha$ tends to $\eta_{\rm t} k_{\rm m}$, i.e.\
\EQ
\alpha\to\eta_{\rm t}\tilde{k}_{\rm m}\quad
\mbox{(for $\meanJJ\neq0$ and $\meanJJ\parallel\meanBB$)}.
\EN
Compared with the kinematic estimate,
$\alpha_{\rm K}\approx\eta_{\rm t}k_{\rm f}$,
$\alpha$ is only quenched by the modified scale separation ratio.
More importantly, $\alpha$ is quenched to a value that is just
slightly above the critical value for the onset of dynamo action,
$\alpha_{\rm crit}=\eta_{\rm T}\tilde{k}_{\rm m}$.
How is it then possible that the fit formula \eq{alpetaCatastrophe}
for $\alpha$ {\it and} $\eta_{\rm t}$ produced reasonable agreement
with the simulations?
The reason is that in the simple case of an $\alpha^2$ dynamo
the solutions are degenerate in the sense that $\meanJJ$
and $\meanBB$ are parallel to each other.
Therefore, the term $\bra{\meanJJ\cdot\meanBB}\meanBB$ is the same as
$\bra{\meanBB^2}\meanJJ$, which means that in the mean EMF the term
$\alpha\meanBB$, where $\alpha$ is given by \Eq{AlphaStationary},
has a component that can be expressed as being parallel to $\meanJJ$.
In other words, the roles of turbulent diffusion (proportional to
$\meanJJ$) and $\alpha$ effect (proportional to $\meanBB$) cannot
be disentangled.
This is the {\it force-free degeneracy} of $\alpha^2$ dynamos
in a periodic box (BB02).
This degeneracy is also the reason why for $\alpha^2$ dynamos the late
saturation behavior can also be described by an algebraic
(non-dynamical, but catastrophic)
quenching formula proportional to $1/(1+R_{\rm m}\bra{\meanBB^2})$
for {\it both} $\alpha$ and $\eta_{\rm t}$, as was done in B01.
To see this, substitute the steady state quenching expression
for $\alpha$, from \Eq{AlphaStationary}, into the expression for $\meanEMF$.
We find
\EQA
\meanEMF=\alpha\meanBB-(\eta+\eta_{\rm t})\meanJJ
={\alpha_{\rm K}
+R_{\rm m}\eta_{\rm t}\bra{\meanJJ\cdot\meanBB}/B_{\rm eq}^2
\over1+R_{\rm m}\bra{\meanBB^2}/B_{\rm eq}^2}\,\meanBB
-\eta_{\rm t}\meanJJ
\nonumber\\
={\alpha_{\rm K}\meanBB
\over 1+R_{\rm m}\bra{\meanBB^2}/B_{\rm eq}^2}
-{\eta_{\rm t}\meanJJ
\over 1+R_{\rm m}\bra{\meanBB^2}/B_{\rm eq}^2},
\label{bothquenched12}
\ENA
which shows that in the force-free case the adiabatic approximation,
together with constant (unquenched) turbulent magnetic diffusivity, becomes
equal to the pair of expressions where both $\alpha$ and $\eta_{\rm t}$
are catastrophically quenched.
This force-free degeneracy is lifted in cases with shear or when the
large scale field is
no longer fully helical (e.g.\ in a nonperiodic domain, and in particular
in the presence of open boundaries).

\subsection{The Keinigs relation and its relevance}

Applying \Eq{dabmdt} to the steady state using
$\meanEMF=\alpha\meanBB-\eta_{\rm t}\mu_0\meanJJ$
(and retaining $\mu_0$ factor), we get
\EQ
\alpha=-\eta\mu_0{\bra{\jj\cdot\bb}\over\bra{\meanBB^2}}
+\eta_{\rm t}k_{\rm m}\quad\mbox{(for periodic domain)},
\label{dabmdt_keinigs}
\EN
where we have defined an effective wavenumber of the large scale
field, $k_{\rm m}=\mu_0\bra{\meanJJ\cdot\meanBB}/\bra{\meanBB^2}$;
see \Eq{JmBmjbm}.
This relation applies only to a closed or periodic box, because
otherwise there would be boundary terms.
Moreover, if the mean field is defined as a volume average, i.e.\
$\meanBB=\bra{\BB}\equiv\BB_0$, then $\mu_0\meanJJ=\nab\times\BB_0=0$,
so $k_{\rm m}=0$ and one has simply
\EQ
\alpha=-\eta\mu_0{\bra{\jj\cdot\bb}\over\bra{\meanBB^2}}
\quad\mbox{(for imposed field)}.
\EN
This equation is due to Keinigs (1983).
For the more general case with $k_{\rm m}\neq0$ this equation has been
discussed in more detail by Brandenburg \& Sokoloff (2002) and
Brandenburg \& Matthaeus (2004).

Let us now discuss the significance of this relation relative to
\Eq{AlphaStationary}.
Both equations apply only in the strictly steady state, of course.
Since we have assumed stationarity, we can replace $\bra{\jj\cdot\bb}$
by $-\bra{\meanJJ\cdot\meanBB}$; see \Eq{JmBmjbm}.
Thus, \Eq{dabmdt_keinigs} reduces to
\EQ
\alpha=\eta_{\rm T}k_{\rm m}
\EN
where $\eta_{\rm T}=\eta+\eta_{\rm t}$ is the total (turbulent and
microscopic) magnetic diffusivity.
This relation is just the condition for a marginally excited dynamo;
see \Eq{a2disperrel}, so it does {\it not} produce any independent
estimate for the value of $\alpha$.
In particular, it does not provide a means of independently testing
\Eq{AlphaStationary}.
The two can however be used to calculate the mean field energy in the
saturated state and we find (BB02)
\EQ
{\bra{\meanBB^2}\over B_{\rm eq}^2}=
{\alpha-\eta_{\rm T}k_{\rm m}\over\eta_{\rm t}k_{\rm m}}
\quad\mbox{(periodic domain)}.
\EN
By replacing $k_{\rm m}$ by an effective value $\tilde{k}_{\rm m}$,
this equation can be generalized to apply also to the case with shear
(for details see BB02).

\subsection{Blackman's multi-scale model: application to helical
turbulence with imposed field}

The restriction to a two scale model may in some cases turn out to be
insufficient to capture the variety of scales involved in astrophysical
bodies.
This is already important in the kinematic stage when the small scale
dynamo obeys the Kazantsev (1968) scaling with a $k^{3/2}$ spectrum
that peaks at the resistive scale.
As the dynamo saturates, the peak moves to the forcing scale.
This lead Blackman (2003) to develop a four scale model where he includes,
in addition to the wavenumbers of the mean field $k_{\rm m}$
($\equiv k_1$) and the
wavenumber of the energy carrying scale of the velocity fluctuations
$k_{\rm f}$ ($\equiv k_2$), also the viscous wavenumber $k_\nu$ ($\equiv k_3$)
and the resistive wavenumber $k_\eta$ ($\equiv k_4$).
The set of helicity equations for the four different scales is
\EQA
\left(\partial_t+2\eta k_1^2\right)H_1
&=&2\bra{\meanEMF_1\cdot\meanBB_1}+2\bra{\meanEMF_2\cdot\meanBB_1},
\\
\left(\partial_t+2\eta k_2^2\right)H_2
&=&-2\bra{\meanEMF_1\cdot\meanBB_1}+2\bra{\meanEMF_2\cdot\meanBB_2},
\\
\left(\partial_t+2\eta k_3^2\right)H_3
&=&-2\bra{\meanEMF_2\cdot\meanBB_1}-2\bra{\meanEMF_2\cdot\meanBB_2},
\\
\left(\partial_t+2\eta k_4^2\right)H_4
&=&0,
\ENA
where $\meanEMF_1$ is the usual electromotive force based on kinetic
helicity at the forcing scale, $k_2$, with feedback proportional to
$H_2$, and $\meanEMF_2$ has no kinetic helicity input but only reacts
to the automatically generated magnetic helicity $H_3$ produced at
the viscous scale $k_3$.
These equations are constructed such that
\EQ
{\partial\over\partial t}\sum_{i=1}^4 H_i=-2\eta\sum_{i=1}^4 k_i^2H_i,
\EN
which is consistent with the magnetic helicity equation \eq{dABmdt}
for the total field.
An important outcome of this model is that in the limit of large $R_{\rm m}$
the magnetic peak travels from $k_3$ to $k_2$ on a dynamical timescale,
i.e.\ a timescale that is independent of $R_{\rm m}$.

\begin{figure}[t!]
\centering\includegraphics[width=0.95\textwidth]{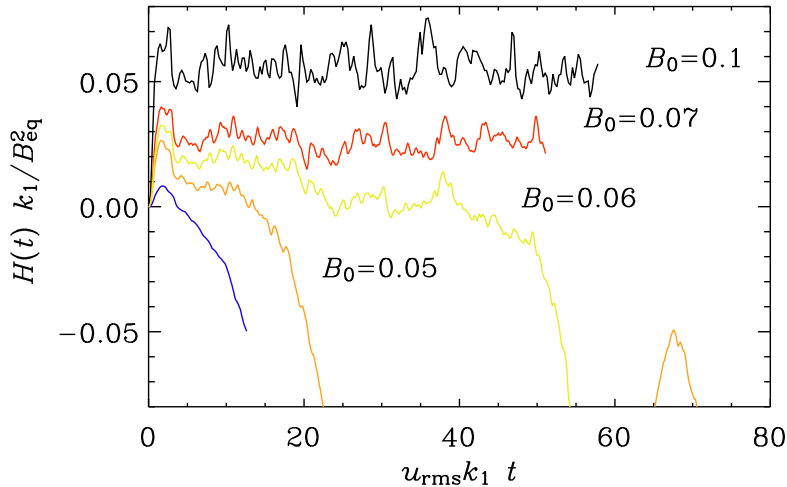}\caption{
Evolution of the total magnetic helicity, $H=H_1+H_{\rm f}$,
as a function of $t$ for different
values of $B_0$, as obtained from the three-dimensional simulation.
Note the change of sign at $B_0\approx B_*\approx0.07$.
[Adapted from Brandenburg \& Matthaeus (2004).]
}\label{Fphelevol}\end{figure}

Brandenburg \& Matthaeus (2004) have applied the general idea to the
case of a model with an applied field.
Here the new scale is the scale of the applied field, but since this
scale is infinite, this field is fixed and not itself subject to an
evolution equation.
Nevertheless, the electromotive force from this field acts as a sink
on the next smaller scale with wavenumber $k_1$, which is the largest
wavenumber in the domain of the simulation.
They thus arrive at the following set of evolution equations,
\EQA
\Big[
\left(\partial_t+2\eta k_0^2\right)H_0
&=& ...+2\bra{\meanEMF_0\cdot\meanBB_0},
\Big]
\\
\left(\partial_t+2\eta k_1^2\right)H_1
&=&-2\bra{\meanEMF_0\cdot\meanBB_0}+2\bra{\meanEMF_1\cdot\meanBB_1},
\label{dotH1}
\\
\left(\partial_t+2\eta k_2^2\right)H_2
&=&-2\bra{\meanEMF_1\cdot\meanBB_1}.
\label{dotHf}
\ENA
The square brackets around the first equation indicate that this
equation is not explicitly included.
From the second equation \eq{dotH1} one can see that there is a competition
between two opposing effects:
the $\alpha$ effect operating on the imposed field $\BB_0$ and the
$\alpha$ effect operating on the $\BB_1$ field on the scale of the box.
When the imposed field exceeds a certain field strength,
$B_0>B_*$, the former will dominate, reversing the sign of the
magnetic helicity at wavenumber $k_1$.
This is actually seen in the simulations of helically forced turbulence
with an imposed field $\BB_0$; see \Fig{Fphelevol}.
We return to this at the end of this section.

The work of Brandenburg \& Matthaeus (2004) was motivated by earlier work of
Montgomery et al.\ (2002) and Milano et al.\ (2003) who showed that,
if the imposed magnetic field is weak or absent, there is a strong nonlocal
transfer of magnetic helicity and magnetic energy from the forcing scale
to larger scales. This leads eventually to the accumulation of magnetic
energy at the scale of the box
(Meneguzzi et al.\ 1981, Balsara \& Pouquet 1999, B01).
As the strength
of the imposed field (wavenumber $k=0$) is increased, the accumulation
of magnetic energy at the scale of the box ($k=1$) becomes more
and more suppressed (Montgomery et al.\ 2002).

In order to solve the model equations, we have to make some assumptions
about the electromotive force operating at $k_0$ and $k_1$.
The large scale magnetic helicity production from the $\alpha$ effect
operating on the imposed field is $\emf_0\cdot\BB_0=\alpha_1\BB_0^2$.
On the other hand, $\emf_1$ at wavenumber $k_1$ is given by
\begin{equation}
\emf_1=\alpha_{\rm f}\BB_1-\eta_{\rm t}\mu_0\JJ_1.
\label{emf1}
\end{equation}
To calculate $\bra{\emf_1\cdot\BB_1}$ in \Eqs{dotH1}{dotHf} we dot
\Eq{emf1} with $\BB_1$, volume average, and note that
$\mu_0\bra{\JJ_1\cdot\BB_1}=k_1^2H_1$ and $\bra{\BB_1^2}=k_1|H_1|$.
The latter relation assumes that the field at wavenumber $k_1$ is fully
helical, but that it can have either sign.
Thus, we have
\begin{equation}
\bra{\emf_1\cdot\BB_1}=\alpha_{\rm f} k_1|H_1|-\eta_{\rm t}k_1^2 H_1.
\end{equation}
The $\alpha$ effects on the two scales are
proportional to the residual magnetic helicity
of Pouquet et al.\ (1976); see \Eq{alpPFM}.
In terms of $H_1$ and $H_2\equiv H_{\rm f}$ we write
\begin{equation}
\alpha_1=\alpha_{\rm K}+\onethird\tau k_1^2 H_1,
\end{equation}
\begin{equation}
\alpha_2=\alpha_{\rm K}+\onethird\tau k_2^2 H_2,
\end{equation}
for the $\alpha$ effect with feedback from $H_1$ and $H_2$,
respectively.

For finite values of $B_0$, the final value of $H_1$ is particularly
sensitive to the value of $\alpha_{\rm K}$ and turns out to be too
large compared with the simulations.
This disagreement with simulations
is readily removed by taking into account that
$\alpha_{\rm K}=-\onethird\tau\bra{\oo\cdot\uu}$ should itself be quenched
when $B_0$ becomes comparable to $B_{\rm eq}$.
Thus, we write
\begin{equation}
\alpha_{\rm K}=\alpha_{\rm K0}/(1+B_0^2/B_{\rm eq}^2),
\end{equation}
which is a good approximation to more elaborate expressions
(R\"udiger \& Kitchatinov 1993).
We emphasize that this equation only applies to $\alpha_{\rm K}$ and is
therefore distinct from Eqs.~\eq{alpetaCatastrophe}, \eq{fullset2},
or \eq{AlphaStationary}.

\begin{figure}[t!]
\centering\includegraphics[width=0.95\textwidth]{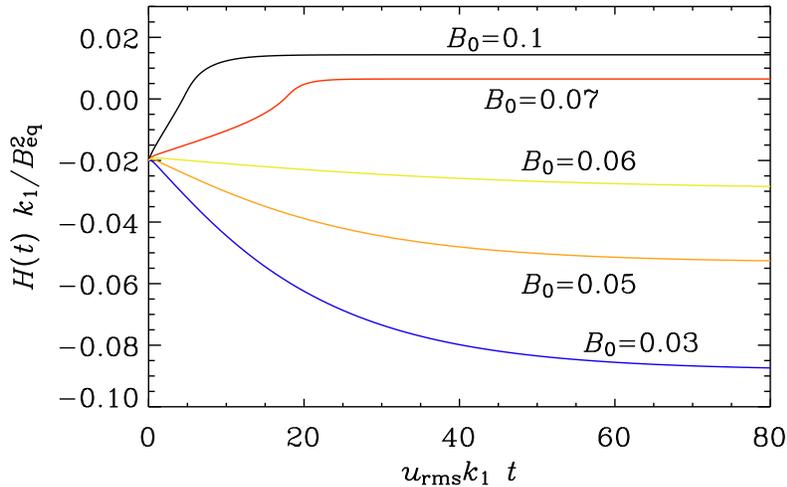}\caption{
Evolution of magnetic helicity as a function of $t$ for different
values of $B_0$, as obtained from the two-scale model.
[Adapted from Brandenburg \& Matthaeus (2004).]
}\label{Fpall}\end{figure}

Under the assumption that the turbulence is fully helical,
the critical value $B_*$ of the imposed field can be estimated by
balancing the two terms on the right hand side of
\Eq{dotHf} and by approximating,
$\alpha\approx\eta_{\rm t}k_{\rm f}$
and $\bra{\jj\cdot\bb}\approx k_{\rm f}B_{\rm eq}^2$.
This yields
\begin{equation}
B_*^2/B_{\rm eq}^2\approx\eta/\eta_{\rm t}\equiv R_{\rm m}^{-1},
\label{critical_Bstar}
\end{equation}
where the last equality is to be understood as
a definition of the magnetic Reynolds number, see also BB02.
For $B_0>B_*$ the sign of the magnetic helicity is the same both at $k=1$
and at $k=k_{\rm f}$, while for $B_0<B_*$ the signs are opposite.

In \Fig{Fpall} we show the result of a numerical integration of
\Eqs{dotH1}{dotHf}.
Both the three-dimensional simulation and the two-scale model show a
similar value of $B_0\approx0.06...0.07$, above which $H_1$ changes sign.
This confirms the validity of our estimate of the critical value $B_*$
obtained from \Eq{critical_Bstar}.
Secondly, the time evolution is slow when $B_0<B_*$ and faster when
$B_0>B_*$.
In the simulation, however, the field attains its final level
for $B_0>B_*$ almost instantaneously, which is not the case in the model.
The significance of this discrepancy remains unclear.
Nevertheless, the level of agreement between the simulations and
3-scale model is surprising, suggesting that the approach can indeed
be quite useful.

\subsection{Alpha effect with open boundaries and shear}
\label{alphaOPEN}

In a recent paper, Brandenburg \& Sandin (2004) have carried out a range
of simulations for different values of the
magnetic Reynolds number, $R_{\rm m}=u_{\rm rms}/(\eta k_{\rm f})$,
for both open and closed boundary conditions using the geometry
depicted on the right hand panel of \Fig{sketch1}.
In order to measure $\alpha$, a uniform magnetic field,
$\BB_0=\const$, is imposed, and the magnetic field is now written as
$\BB=\BB_0+\nab\times\AAA$.
They have determined $\alpha$ by measuring the turbulent
electromotive force, and hence $\alpha=\bra{\emf}\cdot\BB_0/B_0^2$.
Similar investigations have been done before both for forced turbulence
(Cattaneo \& Hughes 1996, B01) and for convective turbulence 
(Brandenburg et al.\ 1990, Ossendrijver et al.\ 2001).

As expected, $\alpha$ is negative when the helicity of the forcing is
positive, and $\alpha$ changes sign when the helicity of the forcing
changes sign.
The magnitudes of $\alpha$ are however different in the two cases:
$|\alpha|$ is larger when the helicity of the forcing is negative.
In the sun, this corresponds to the sign of helicity in the northern
hemisphere in the upper parts of the convection zone.
This is here the relevant case, because the differential rotation
pattern of the present model also corresponds to the northern hemisphere. 

\begin{figure}[t!]
\centering\includegraphics[width=0.95\textwidth]{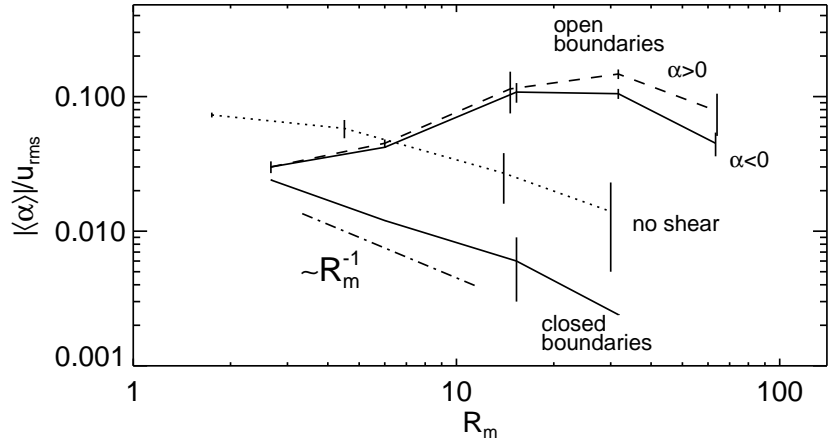}\caption{
Dependence of $|\bra{\alpha}|/u_{\rm rms}$ on $R_{\rm m}$
for open and closed boundaries.
The case with open boundaries and negative helicity is shown as a dashed line.
Note that for $R_{\rm m}\approx30$ the $\alpha$ effect
is about 30 times smaller when the boundaries are closed.
The dotted line gives the result with open boundaries but no shear.
The vertical lines indicate the range obtained by calculating
$\alpha$ using only the first and second half of the time interval.
[Adapted from Brandenburg \& Sandin (2004).]
}\label{palp_sum}\end{figure}

There is a striking difference between the cases with open and
closed boundaries which becomes particularly clear when comparing
the averaged values of $\alpha$ for different magnetic Reynolds
numbers; see \Fig{palp_sum}.
With closed boundaries $\alpha$ tends to zero like $R_{\rm m}^{-1}$,
while with open boundaries $\alpha$ shows no such decline.
There is also a clear difference between the cases with and without shear
together with open boundaries in both cases.
In the absence of shear (dashed line in \Fig{palp_sum}) $\alpha$ declines
with increasing $R_{\rm m}$, even though for small values of $R_{\rm m}$
it is larger than with shear.
The difference between open and closed boundaries
will now be discussed in terms of a current helicity
flux through the two open open boundaries of the domain.

\subsection{Current helicity flux}

It is suggestive to interpret the above results in terms of the
dynamical $\alpha$ quenching model.
However, \Eq{fullset2} has to be generalized to take the divergence
of the flux into account.
In order to avoid problems with the gauge, it is advantageous to
work directly with $\overline{\jj\cdot\bb}$ instead of
$\overline{\aaaa\cdot\bb}$.
Using the evolution equation, $\partial\bb/\partial t=-\nab\times\ee$,
for the fluctuating magnetic field, where $\ee=\EE-\meanEE$ is the
small scale electric field and $\meanEE=\eta\mu_0\meanJJ-\meanEMF$ the
mean electric field, one can derive the equation
\begin{equation}
{\partial\over\partial t}\overline{\jj\cdot\bb}
=-2\,\overline{\ee\cdot\cc}-\nab\cdot\meanFF_C^{\rm SS},
\label{jc_evolution}
\end{equation}
where
\begin{equation}
\meanFF_C^{\rm SS}=\overline{2\ee\times\jj}
+\overline{(\nab\times\ee)\times\bb}/\mu_0,
\label{meanFFc}
\end{equation}
is the current helicity flux from the small scale field, and
$\cc=\nab\times\jj$ the curl of the small scale current density,
$\jj=\JJ-\meanJJ$.
In the isotropic case,
$\overline{\ee\cdot\cc}\approx k_{\rm f}^2\overline{\ee\cdot\bb}$, where
$k_{\rm f}$ is the typical wavenumber of the fluctuations,
here assumed to be the forcing wavenumber.
Ignoring the effect of the mean flow on $\meanEMF$ [as is usually done;
but see Krause \& R\"adler (1980) and the recent on the shear current
effect by Rogachevskii \& Kleeorin (2003, 2004); see \Sec{OpenSurface}],
we obtain
\begin{equation}
\overline{\ee\cdot\bb}
\approx-\overline{(\uu\times\BB_0)\cdot\bb}+\eta\mu_0\overline{\jj\cdot\bb}
=\meanEMF\cdot\meanBB+\eta\mu_0\overline{\jj\cdot\bb},
\end{equation}
where we have used $\overline{\uu\times\bb}=\meanEMF$ and
$\BB_0=\meanBB$.
Using standard expressions for the turbulent magnetic diffusivity,
$\eta_{\rm t}=\onethird\tau u_{\rm rms}^2$, and the equipartition
field strength, $B_{\rm eq}=\sqrt{\mu_0\rho}\,u_{\rm rms}$,
we eliminate $\tau$ via
\begin{equation}
\onethird\tau=\mu_0\rho_0\eta_{\rm t}/B_{\rm eq}^2.
\end{equation}
This leads to an explicitly time dependent formula for $\alpha$,
\begin{equation}
{\partial\alpha\over\partial t}=-2\eta_{\rm t} k_{\rm f}^2\left(
{\meanEMF\cdot\meanBB
+\half k_{\rm f}^{-2}\nab\cdot\mu_0\meanFF_C^{\rm SS}\over B_{\rm eq}^2}
+{\alpha-\alpha_{\rm K}\over R_{\rm m}}\right).
\label{fullset2flux}
\end{equation}
This equation is similar to that of Kleeorin et al.\ (2000, 2002, 2003)
who considered the flux of magnetic helicity instead of current helicity.

Making use of the adiabatic approximation, i.e.\ putting the rhs of
\Eq{fullset2flux} to zero, one arrives at the algebraic
steady state quenching formula ($\partial\alpha/\partial t=0$)
\begin{equation}
\alpha={\alpha_{\rm K}
+R_{\rm m}\left(\eta_{\rm t}\mu_0\meanJJ\cdot\meanBB
-\half k_{\rm f}^{-2}\nab\cdot\mu_0\meanFF_C^{\rm SS}\right)/B_{\rm eq}^2
\over1+R_{\rm m}\meanBB^2/B_{\rm eq}^2}.
\label{AlphaStationaryFlux}
\end{equation}
In the absence of a mean current, e.g.\ if the mean field is defined
as an average over the whole box, then $\meanBB\equiv\BB_0=\const$,
and $\meanJJ=0$, so \Eq{AlphaStationaryFlux} reduces to
\begin{equation}
\alpha={\alpha_{\rm K}
-\half k_{\rm f}^{-2}R_{\rm m}\nab\cdot\mu_0\meanFF_C^{\rm SS}/B_{\rm eq}^2
\over1+R_{\rm m}\BB_0^2/B_{\rm eq}^2}.
\label{AlphaStationaryFlux_noJ}
\end{equation}
This expression applies to the present case, because we consider
only the statistically steady state and we also define the mean field
as a volume average.

For closed boundaries, $\bra{\nab\cdot\meanFF_C^{\rm SS}}=0$, and so
\Eq{AlphaStationaryFlux_noJ} clearly reduces to a catastrophic quenching
formula, i.e.\ $\alpha$ vanishes in the limit of large magnetic Reynolds
numbers as
\begin{equation}
\alpha^{\rm(closed)}={\alpha_{\rm K}
\over1+R_{\rm m}\BB_0^2/B_{\rm eq}^2}\to R_{\rm m}^{-1}
\quad\mbox{(for $R_{\rm m}\to\infty$)}.
\label{AlphaStationary_noFlux}
\end{equation}
The $R_{\rm m}^{-1}$ dependence is confirmed by the simulations
(compare with the dash-dotted line in \Fig{palp_sum}).
On the other hand, for open boundaries the limit $R_{\rm m}\to\infty$
gives
\begin{equation}
\alpha^{\rm(open)}\to
-(\nab\cdot\mu_0\meanFF_C^{\rm SS})/(2k_{\rm f}^2\BB_0^2)
\quad\mbox{(for $R_{\rm m}\to\infty$)},
\label{AlphaOpenLimit}
\end{equation}
which shows that losses of negative helicity, as observed in the northern
hemisphere of the sun, would enhance a positive $\alpha$ effect
(Kleeorin et al.\ 2000).
In the simulations, the current helicity flux is found to be independent
of the magnetic Reynolds number.
This explains why the $\alpha$ effect no longer shows the catastrophic
$R_{\rm m}^{-1}$ dependence (see \Fig{palp_sum}).
In principle it is even conceivable that with $\alpha_{\rm K}=0$
a current helicity flux can be generated, for example by shear,
and that this flux divergence could drive a dynamo, as was suggested
by Vishniac \& Cho (2001).
It is clear, however, that for finite values of $R_{\rm m}$ this would
be a non-kinematic effect requiring the presence of an already finite
field (at least of the order of $B_{\rm eq}/R_{\rm m}^{1/2}$).
This is because of the $1+R_{\rm m}\BB_0^2/B_{\rm eq}^2$ term in the
denominator of \Eq{AlphaStationaryFlux_noJ}.
At the moment we cannot say whether this is perhaps the effect leading
to the nonhelically forced turbulent dynamo discussed in \Sec{OpenSurface},
or whether it is perhaps the $\ddelta\times\meanJJ$ or shear-current
effect that was also mentioned in that section.

\section{What about $\eta$ quenching?}

As we have seen above, in a closed domain the value of $\alpha$ in the
saturated state cannot conclusively be determined without also determining
at the same time the turbulent magnetic diffusivity.
There are different ways of determining $\eta_{\rm t}$.
The values are not necessarily all in agreement with each other,
because one does not know whether the mean field equation, where
$\eta_{\rm t}$ enters, is correct and applicable.
We report here a few different examples where $\eta_{\rm t}$ has been determined.

\subsection{Direct measurements in a working dynamo}

We first consider the case of a helical turbulent dynamo without shear (B01)
and compare it with a simple mean-field $\alpha^2$ dynamo.
Assuming that $\alpha$ is uniform, we can use \Eq{constcoeff} and,
assuming furthermore
that $\alpha<0$ (which is the case when the helicity of the forcing
is positive, as in B01), the solution is
\EQ
\meanBB=(b_x\cos k_1z, b_y\sin k_1z, 0)^T.
\EN
The time-dependent equations can then be written as
\EQ
{\dd b_x\over\dd t}=|\alpha|b_y-\eta_{\rm T}k_1^2b_x,
\EN
\EQ
{\dd b_y\over\dd t}=|\alpha|b_x-\eta_{\rm T}k_1^2b_y.
\EN
In an isotropic, homogeneous $\alpha^2$ dynamo, the eigenfunction
obeys $b_x=b_y$.

\begin{figure}[t!]
\centering\includegraphics[width=0.8\textwidth]{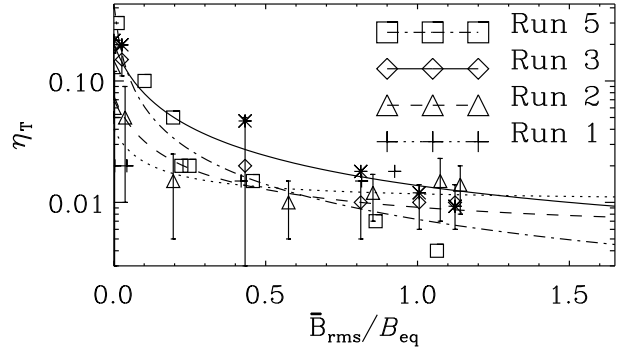}\caption{
Result for $\eta_{\rm T}$ for different values of $R_{\rm m}$.
The lines represent the fits described in the text. In the plot of
$\eta_{\rm T}$ the asterisks denote $|\alpha|-\lambda$ for the
$R_{\rm m,forc}=120$ run, which agrees reasonably well with $\eta_{\rm T}$.
[Adapted from B01.]
}\label{Fpalpb_sep2}\end{figure}

We now assume that, at some particular time, we put $b_x=0$, for example.
This means that $b_x(t)$ will first grow linearly in time at a rate
that is proportional to $\alpha$ like $b_x\approx|\alpha|k_1 b_y$.
At the same time as $b_x$ grows, $b_y$ will first decrease at a
rate that is proportional to $\eta_{\rm T}$.
This allows an independent estimate of $b_x$ and $b_y$
by solving the matrix equation
\EQ
\pmatrix{k_1b_y & -k_1^2b_x\cr k_1 b_x & -k_1^2 b_y}
\pmatrix{|\alpha|\cr\eta_{\rm T}}
=\pmatrix{\dd b_x/\dd t\cr\dd b_y/\dd t}.
\EN
The result for $\alpha$ is found to be roughly consistent with that of
Cattaneo \& Hughes (1996), and the result for $\eta_{\rm T}$ is reproduced in
\Fig{Fpalpb_sep2}, and can be described by the fit formula
\EQ
\eta_{\rm t}={\eta_{\rm t0}\over1+\tilde{g}|\meanBB|/B_{\rm eq}}
\label{fit_etat}
\EN
with $\tilde{g}\approx16$.
This expression needs to be compared with that obtained from
other approaches.

The fact that the results obtained for $\alpha$ by using this
approach are consistent with that for a uniform field is quite surprising
and unexpected.
This agreement probably indicates that in this type of simulation
$\alpha$ is independent of scale -- at least in the scale range corresponding
to wavenumbers $k=k_1$ ($=1$) and $k=0$.
In general, this may not be true.
Indeed, in the case of accretion discs some numerical evidence for scale
dependence of $\alpha$ and $\eta_{\rm t}$ has been found
(Brandenburg \& Sokoloff 2002).

\subsection{Measurements in an $\alpha\Omega$ dynamo}

In the case of an $\alpha\Omega$ dynamo the cycle frequency
$\omega_{\rm cyc}$ depends directly on the nonlinearly suppressed
value of $\eta_{\rm T}$
\EQ
\omega_{\rm cyc}=\eta_{\rm T}k_1^2
\quad\mbox{($\eta_{\rm T}$ is quenched)},
\label{ocycQUENCHED}
\EN
see BB02 (their Sect.~4.2).
The estimates of BBS01 indicated that the dynamo numbers based on shear,
$C_\Omega=S/(\eta_{\rm T}k_1^2)$, is between 40 and
80, whilst the total dynamo number (${\cal D}=C_\alpha C_\Omega$) is between
10 and 20 (see BBS01),
and hence $C_\alpha=\alpha/(\eta_{\rm T}k_1)\approx0.25$.
Thus, shear dominates strongly
over the $\alpha$ effect ($C_\Omega/C_\alpha$ is between 150 and 300), which
is typical for $\alpha\Omega$-type behavior (i.e.~oscillations) rather
than $\alpha^2$-type
behavior which would start when $C_\Omega/C_\alpha$ is below about 10
(e.g.\ Roberts \& Stix 1972).

The results shown in \Tab{Tao2}
suggest that the period in this oscillatory dynamo
is controlled by the microscopic magnetic diffusivity,
because $\omega_{\rm cyc}/(\eta k_1^2)$ is approximately independent
of $R_{\rm m}$.
Using \Eq{ocycQUENCHED}, this means that
$\eta_{\rm T}({\rm quenched})/\eta={\cal O}(10)$ for
$R_{\rm m}$ between 30 and 200.
This result would favor a model where $\eta_{\rm T}$ is still
quenched in an $R_{\rm m}$-dependent fashion.
In the next section we show that the apparent
$R_{\rm m}$-dependent $\eta_{\rm t}$ quenching can
easily also be produced when the field possesses a helical component.

\begin{table}[t!]\caption{
Summary of the main properties of the three-dimensional
simulations with shear. Here, $\eta/(c_{\rm s}k_1)$ is
the magnetic diffusivity in units of the sounds speed and
the wavenumber of the domain,
and $\omega_{\rm cyc}=2\pi/T_{\rm cyc}$ is the
cycle frequency. In Run~(iii) there is no clear cycle visible.
[Adapted from BDS02.]
}\vspace{12pt}\centerline{\begin{tabular}{lccc}
\hline
Run &  (i) & (ii) & (iii) \\
\hline
$\eta/(c_{\rm s}k_1)$ & $10^{-3}$ & $5\times10^{-4}$ & $2\times10^{-4}$ \\
$\nu/\eta$ & 5 & 10 & 25 \\
$R_{\rm m}=\bra{\uu^2}^{1/2}/(\eta k_1)$ & 30 & 80 & 200 \\
$C_\Omega=\bra{\meanUU^2}^{1/2}/(\eta k_1)$ & 1000 & 2000 & 4000 \\
$\bra{\bb^2}/B_{\rm eq}^2$ &  4 & 6 & 20 \\
$\bra{\meanBB^2}/B_{\rm eq}^2$ & 20 & 30 & 60 \\
$\epsilon_{\rm m}=\mu_0\bra{\meanJJ\cdot\meanBB}/\bra{\meanBB^2}$ & 0.11 & 0.06 & 0.014 \\
$\omega_{\rm cyc}/(\eta k_1^2)$ &  8\ldots9 & 6\ldots12 & $\ge10?$ \\
\hline
\label{Tao2}\end{tabular}}\end{table}

Looking at the scaling of the cycle frequency with resistivity may
be quite misleading in the present case, because the large scale
magnetic field exceeds the kinetic energy by a large factor (20--30).
This would always lead to the usual (non-catastrophic) quenching of
$\alpha$ and $\eta_{\rm t}$.
Furthermore, such strong magnetic fields will affect the mean shear flow.
Most important is perhaps the fact that in the simulation of BBS01
the shear flow varies sinusoidally in the cross stream direction, so
the mean field depends on the two coordinate directions perpendicular
to the streamwise direction.
For this reason BB02 solved the mean field and dynamical quenching
equations in a 2-dimensional model.
It turned out to be important to allow for non-catastrophic
quenching of $\eta_{\rm t}$ using \Eq{fit_etat}
where the value of $\tilde{g}$ has been varied between 0 and 3.
The asymptotic $1/B$ behavior (as opposed to $1/B^2$, for example) was
motivated both by simulations (B01) and analytic results
(Kitchatinov et al.\ 1994, Rogachevskii \& Kleeorin 2001).

In order to see whether the models can be made to match the direct
simulations, several input parameters were varied.
It should be kept in mind, however, that not all input parameters
are well known.
This has to do with the uncertainty in the correspondence between
the magnetic Reynolds number in the model (which measures
$\eta_{\rm t0}/\eta$) and the simulations [where it is defined
as $u_{\rm rms}/(\eta k_{\rm f})$].
Likewise, the dynamo number $C_\alpha=\alpha/(\eta_{\rm T} k_1)$
is not well determined.
Nevertheless, many of the output parameters are reasonably well
reproduced; see \Tab{Tao3}.

\begin{table}[t!]\caption{
Results from the simulations of BBS01 and BDS02,
compared with those of 2-dimensional mean field models.
Model results that are in fair agreement with the simulations
are highlighted in bold face.
Here, $Q$ is the ratio of toroidal to poloidal rms field.
}\vspace{12pt}\centerline{\begin{tabular}{cccrcccccccc}
\hline
Model
& ~~$R_{\rm m}$~~
& ~$C_\alpha$~
& ~$C_\Omega$~
& ~~$\tilde{g}$~~
& ~~${\displaystyle{S\over\eta k_1^2}}$~~
& ~~${\displaystyle{\bra{\bb^2}\over B_{\rm eq}^2}}$~~
& ~~${\displaystyle{\bra{\meanBB^2}\over B_{\rm eq}^2}}$~~
& ~~$Q^{-1}$~~
& ~~$\epsilon_{\rm m}$~~
& ~~${\displaystyle{\omega_{\rm cyc}\over S}}$~~
& ~~${\displaystyle{\lambda\over S}}$~~ \\
\hline
BBS01&$80$&1--2&--&--&2000& 6  & 30  & 0.014 & 0.06 &0.008&0.015\\
R1   &$20$&1.0 &100&0&2000&0.20&{\bf15}& 0.031 &{\bf}0.065&0.016&0.044 \\
AG2  &$100$&0.5 & 20&3&2000&0.10&{\bf22}&{\bf}0.011 & 0.024&{\bf0.006}&{\bf0.021}\\
\hline
BDS02&$30$&1--2&--&--&1000& 4  & 20  & 0.018 & 0.11 &0.014&0.006 \\
s3   &$30$&0.35& 33&1&1000&0.07&  6  & 0.029 &{\bf0.061}&{\bf0.014}&{\bf0.016}\\
S1   &$30$&0.35& 33&3&1000&0.07&{\bf19}& 0.009 & 0.019&0.005&0.016 \\
\hline
\label{Tao3}\end{tabular}}\end{table}

\subsection{Decay experiments}

\begin{figure}[t!]
\centering\includegraphics[width=0.7\textwidth]{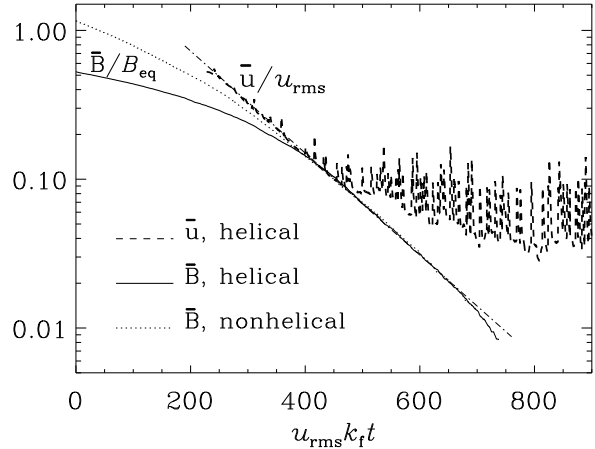}\caption{
Decay of large scale helical velocity and magnetic fields
(dashed and solid lines, respectively).
The graph of $\meanUU(t)$ has been shifted so that both $\meanUU(t)$
and $\meanB(t)$ share the same tangent (dash-dotted line), whose
slope corresponds to
$\nu_{\rm t}=\eta_{\rm t}=0.86u_{\rm rms}/k_{\rm f}$.
The decay of a nonhelical magnetic field is shown for
comparison (dotted line).
[Adapted from Yousef et al.\ (2003)]
}\label{Fpn_comp_prm}\end{figure}

Finally, we consider the decay of a magnetic field.
This provides a fairly straightforward method of determining
$\eta_{\rm T}$ from the decay rate $\lambda$ of a sinusoidal
field with wavenumber $k_1$, so $\lambda=\eta_{\rm T}k_1^2$.
The result reported by Yousef et al.\ (2003) suggests that
\EQ
\nu_{\rm t}\approx\eta_{\rm t}=(0.8\ldots0.9)\times u_{\rm rms}/k_{\rm f}
\quad\mbox{(for $\meanBB^2\ll B_{\rm eq}^2$)}.
\label{nut_etat}
\EN
Once the mean flow profile has decreased below a certain level ($<0.1u_{\rm rms}$),
it cannot decay further and continues to fluctuate around $0.08u_{\rm rms}$,
corresponding to the level of the rms velocity of the (forced!) turbulence at $k=k_1$
(see the dashed line in \Fig{Fpn_comp_prm}).

\begin{figure}[t!]
\centering\includegraphics[width=0.99\textwidth]{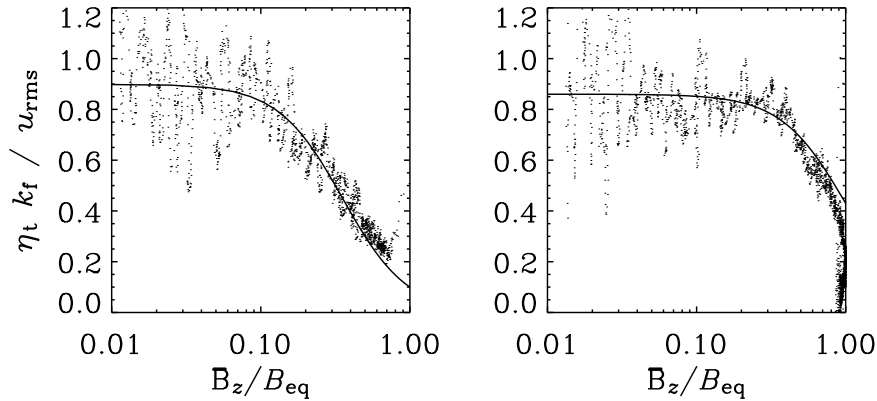}
\caption{
Dependence of the turbulent diffusion coefficient on the magnitude of
the mean field.
$R_{\rm m}\approx20$.
Left: The initial field is helical and corresponds to data
points on the right hand side of the plot. 
The data are best fitted by $\tilde{g}=8=0.4R_{\rm m}$.
Right: the same for the nonhelical case.
The data are best fitted by $a=1$, independent of $R_{\rm m}$.
[Adapted from Yousef et al.\ (2003)]
}\label{Fpdecay_law}\end{figure}

The quenching of the magnetic diffusivity,
$\eta_{\rm t}=\eta_{\rm t}(\meanBB)$, can be obtained
from one and the same run by simply determining the decay rate,
$\lambda_B(\meanB)$, at different times, corresponding
to different values of $\meanB=|\meanBB|$.
To describe departures from purely exponential decay one can adopt a
$\meanBB$-dependent $\eta_{\rm t}$ expression of the form \eq{fit_etat}.
It turns out that the value of $\tilde{g}$ is not universal
and depends on the field geometry.
This is easily demonstrated by comparing the decay of helical and
nonhelical initial fields; see \Fig{Fpdecay_law}.

In the next section we show that the slower decay of $\meanBB$, and
hence the implied stronger quenching of $\eta_{\rm t}$, can also be
described by a self-induced magnetic $\alpha$ effect which acts such
as to decrease the decay rate.
In the case of a helical initial field, we have
$\meanJJ\times\meanBB=0$, i.e.\ the large scale field is force-free
and interacts only weakly with the turbulence.

Thus, the indications here are that for non-helical fields,
$\eta_{\rm t}$ is {\it not} catastrophically quenched.
A resistively slow decay rate occurs however when the magnetic
field is helical, but this is {\it not} to be explained by a
catastrophically quenched $\eta_{\rm t}$, but by the
magnetic $\alpha$ effect, $\alpha_{\rm M}$, that tries to keep
the magnetic field as large as possible, just as enforced by the
magnetic helicity constraint.
The phenomenon, described in this way, may be more easily described
in terms of helicity conservation, because the system has magnetic
helicity that can only decay slowly on a resistive time scale,
hence lowering the apparent turbulent diffusivity down to the
microscopic value $\eta$.
This will be explained in more detail in the next section.

\subsection{Taylor relaxation or selective decay}
\label{Sdynquench}

In the case of a helical field with
$\meanBB^2/B_{\rm eq}^2\ga R_{\rm m}^{-1}$
the slow decay of $\meanBB$
is related to the conservation of magnetic helicity.
As already discussed by BB02,
this behavior is related to the phenomenon of selective
decay (e.g.\ Montgomery et al.\ 1978) and can be described by the
dynamical quenching model.
This model applies even to the case where the turbulence is nonhelical and
where there is no $\alpha$ effect in the usual sense.
However, the magnetic contribution to $\alpha$ is still
non-vanishing, because it is
driven by the helicity of the large scale field.

To demonstrate this quantitatively, Yousef et al.\ (2003) have
adopted the one-mode approximation ($\kk=\kk_1$) with
$\meanBB=\hat\BB\exp(\ii\kk_1z)$, the mean-field induction equation
\EQ
{\dd\hat\BB\over\dd t}=\ii\kk_1\times\hat{\emf}
-\eta k_1^2\hat\BB,
\label{ODE1}
\EN
together with the dynamical $\alpha$-quenching formula \eq{fullset2},
\EQ
{\dd\alpha\over\dd t}=-2\eta k_{\rm f}^2
\left(\alpha+\tilde{R}_{\rm m}{{\rm Re}(\hat{\emf}^*\cdot\hat\BB)
\over B_{\rm eq}^2}\right),
\label{ODE2}
\EN
where
\EQ
\hat{\emf}=\alpha\hat\BB-\eta_{\rm t}\ii\kk_1\times\hat\BB
\label{hat_emf}
\EN
is the
electromotive force, and $\tilde{R}_{\rm m}$ is defined as the ratio
$\eta_{\rm t0}/\eta$, which is expected to be close to the value of
$R_{\rm m}$.

\begin{figure}[t!]
\centering\includegraphics[width=0.7\textwidth]{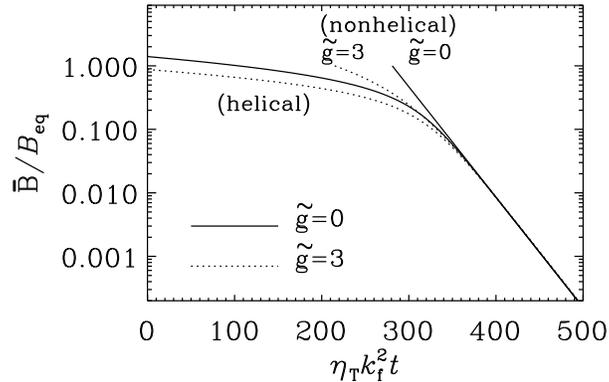}\caption{
Dynamical quenching model with helical and nonhelical initial fields.
The quenching parameters are $\tilde{g}=0$ (solid line) and 3 (dotted line).
The graph for the nonhelical cases has been shifted in $t$ so that one
sees that the decay rates are asymptotically equal at late times.
The value of $\eta_{\rm T}$ used to normalize the abscissa is based
on the unquenched value.
[Adapted from Yousef et al.\ (2003)]
}\label{Fpselected_decay}\end{figure}

In \Fig{Fpselected_decay} we show the evolution of $\meanBB/B_{\rm eq}$
for helical and nonhelical initial conditions, $\hat\BB\propto(1,\ii,0)$
and $\hat\BB\propto(1,0,0)$, respectively.
In the case of a nonhelical field, the decay rate is not quenched at all,
but in the helical case quenching sets in for
$\meanBB^2/B_{\rm eq}^2\ga R_{\rm m}^{-1}$.
In the helical case, the onset of quenching at
$\meanBB^2/B_{\rm eq}^2\approx R_{\rm m}^{-1}$ is well reproduced by
the simulations.
In the nonhelical case, however, some weaker form of quenching sets in
when $\meanBB^2/B_{\rm eq}^2\approx1$
(see the right hand panel of \Fig{Fpdecay_law}).
We refer to this as standard quenching (e.g.\ Kitchatinov et al.\ 1994)
which is known to be always present; see \Eq{fit_etat}.
BB02 found that, for a range of different values of $R_{\rm m}$,
$\tilde{g}=3$ resulted in a good description of
the simulations of cyclic $\alpha\Omega$-type dynamos (BDS02).

Yousef et al.\ (2003) also showed
that the turbulent magnetic Prandtl number
is indeed independent of the microscopic magnetic Prandtl number.
The resulting values of the flow Reynolds number,
$\mbox{Re}=u_{\rm rms}/(\nu k_{\rm f})$, varied between 20 and 150,
giving $P_{\rm m}$ in the range between 0.1 and 1.
Within plot accuracy the three values of $\lambda_B$ turn out
to be identical in the interval where the decay is exponential.

\section{Conclusions}

In the present review we have tried to highlight some of the recent
discoveries that have led to remarkable advances in the theory of
mean field dynamos.
Of particular importance are the detailed confirmations of various aspects
of mean field theory using helically forced turbulence simulations.
The case of homogeneous turbulence with closed or periodic boundary
conditions is now fairly well understood.
In all other cases, however, the flux of current helicity becomes
important.
The closure theory of these fluxes is still a matter of ongoing
research (Kleeorin et al.\ 2000, 2002, 2003),
Vishniac \& Cho (2001), Subramanian \& Brandenburg (2004), and
Brandenburg \& Subramanian (2004).
The helicity flux of Vishniac \& Cho (2001)
has been independently confirmed (Subramanian \& Brandenburg 2004).
A more detailed investigation of current helicity fluxes appears
to be quite important when one tries to get qualitative and
quantitative agreement between simulations and theory.

The presence of current helicity fluxes is particularly important
when there is also shear.
This was already recognized by Vishniac \& Cho (2001) who applied
their calculations to the case of accretion discs where shear is
particularly strong.
In the near future it should be possible to investigate the emergence
of current helicity flux in more detail.
This would be particularly interesting in view of the many observations of
coronal mass ejections that are known to be associated with significant
losses of magnetic helicity and hence also of current helicity
(DeVore 2000, D\'emoulin et al.\ 2002, Gibson et al.\ 2002).

In order to be able to model coronal mass ejections it should be
particularly important to relax the restrictions imposed by the vertical
field conditions employed in the simulations of Brandenburg \& Sandin (2004).
A plausible way of doing this would be to include a simplified version
of a corona with enhanced temperature and hence decreased density,
making this region a low-beta plasma.

In the context of accretion discs the importance of adding a corona
is well recognized (Miller \& Stone 2000), although its influence on
large scale dynamo action is still quite open.
Regarding hydromagnetic turbulence in galaxies, most simulations to date
do not address the question of dynamo action (Korpi et al.\ 1999,
de Avillez \& Mac Low 2002).
This is simply because here the turbulence is driven by supernova
explosions which leads to strong shocks.
These in turn require large numerical diffusion, so the effective
magnetic Reynolds number is probably fairly small and dynamo action
may only be marginally possible.
In nonhelically driven turbulence has been applied to
the galactic medium to argue that it is dominated by small scale fields
(Schekochihin et al.\ 2002), but the relative importance of small scale
fields remains still an open question (Haugen et al.\ 2003).
Galaxies are however rotating and vertically stratified, so the flows
should be helical, but in order to say anything about magnetic helicity
evolution, much larger magnetic Reynolds numbers are necessary.
At the level of mean field theory the importance of magnetic helicity
fluxes is well recognized.
The explicitly time-dependent dynamical $\alpha$ quenching equation
with magnetic helicity fluxes has been included in mean field simulations
(Kleeorin et al.\ 2000, 2002, 2003), but the form of the adopted fluxes
is to be clarified in view of the differences with the results of
Vishniac \& Cho (2001) and Subramanian \& Brandenburg (2004).
Nevertheless, given that the form of the dynamical quenching equations is
likely to be still incomplete, it remains to be demonstrated, using
simulations, that magnetic or current helicity fluxes do really allow
the dynamo to saturate on a dynamical time scale.

\section*{Acknowledgements}
The Danish Center for Scientific Computing is acknowledged for granting
time on the Horseshoe cluster.
This work has been completed while being on sabbatical at the
Isaac Newton Institute for Mathematical Sciences in Cambridge.


\vfill\bigskip\noindent\tiny\begin{verbatim}
$Header: /home/brandenb/CVS/tex/mhd/wielebinski/paper.tex,v 1.57 2004/12/13 16:06:55 brandenb Exp $
\end{verbatim}

\end{document}